\documentclass[journal=jcisd8,manuscript=article,layout=traditional]{achemso}
\setkeys{acs}{articletitle = true}
\setkeys{acs}{etalmode = truncate}
\setkeys{acs}{maxauthors = 0}

\usepackage[version=3]{mhchem} 

\usepackage{xr-hyper}
\makeatletter
\newcommand*{\addFileDependency}[1]{
\typeout{(#1)}
\@addtofilelist{#1}
\IfFileExists{#1}{}{\typeout{No file #1.}}
}\makeatother

\usepackage{siunitx, amsmath, bbold, dsfont, hyperref, multirow, subcaption, graphicx}

\setkeys{acs}{doi = true}
\usepackage{xcolor}
\hypersetup{
    colorlinks,
    linkcolor={teal},
    citecolor={teal},
    urlcolor={teal}
}
\SectionNumbersOn

\usepackage{algorithm, fancyvrb, listings}
\usepackage{cuted}
\usepackage{multirow}
\usepackage{makecell}  

\newcommand{\doi}[1]{\href{http://dx.doi.org/#1}{\nolinkurl{#1}}}
\DeclareSIUnit\angstrom{\text{\normalfont\AA}}
 \DeclareSIUnit\hartree{\mathrm {\ensuremath {E}}_{\mathrm {h}}}
 
\author{Andrea Levy}
\affiliation{Laboratory of Computational Chemistry and Biochemistry, Ecole Polytechnique Fédérale de Lausanne, Lausanne, Switzerland}
\author{Andrej Antal\'{i}k}
\affiliation{Laboratory of Computational Chemistry and Biochemistry, Ecole Polytechnique Fédérale de Lausanne, Lausanne, Switzerland}
\author{J\'{o}gvan Magnus Haugaard Olsen}
\affiliation{DTU Chemistry, Technical University of Denmark, Lyngby, Denmark}
\author{Ursula Rothlisberger}
\affiliation{Laboratory of Computational Chemistry and Biochemistry, Ecole Polytechnique Fédérale de Lausanne, Lausanne, Switzerland}
\email{ursula.roethlisberger@epfl.ch}

\title{Atom-centered electric multipole moments dynamically generated from QM/MM MD simulations}

\makeatletter
\if@twocolumn

\let\maketitle\relax
\fi
\makeatother

\begin{document}







 
\begin{abstract}
Atom-centered electric multipole moments can be extremely useful in chemistry as they enable the systematic mapping of a complex electrostatic problem to a simpler model.
However, since they do not correspond to physical observables, there is no unique way to define them.
In this work, we present an extension of the dynamically generated RESP charges (D-RESP) method, referred to as xDRESP, where atom-centered multipoles are computed from mixed quantum mechanics/molecular mechanics (QM/MM) molecular dynamics simulations.
We compare the ability of xDRESP charges to reproduce the electrostatic potential, as well as molecular multipoles, against the performance of fixed point-charge models commonly used in force fields. 
Moreover, we highlight cases where xDRESP atomic multipoles can provide valuable information about chemical systems, such as indicating when polarization plays a significant role, and chemical reactions, in which xDRESP atomic multipoles can be used as an on-the-fly analysis tool to track changes in electron density.
\end{abstract}

\section{\label{sec:intro}Introduction}
Atomic charges or, more generally, atom-centered electric multipole moments, are an extremely useful concept in chemistry that can help to describe the electrostatic properties of complex molecular systems by using a simple model.
Yet, as they do not correspond to a physical observable, i.e., there is no quantum mechanical operator whose eigenvalues correspond to such quantities \cite{oliferenko2006atomic, cho2020atomic}, they cannot be defined in a unique way in a quantum mechanical framework. 

Over the years, researchers have proposed different methods to define effective atomic charges, which can be classified into three main categories.
The first are methods based on the partitioning of the wavefunction, such as Mulliken \cite{mulliken1955electronic} or Löwdin \cite{lowdin1950charges} charges, natural population analysis \cite{reed1985natural}, or CM\textit{x} charge models \cite{storer1995CM1, li1998CM2, winget2002CM3, kelly2005CM4, olson2007CM4M, marenich2012CM5} (\textit{x}=1--5 or 4M).
Another class is represented by methods that partition the electron density distribution, such as charges from Bader's Atoms-In-Molecules (AIM) analysis \cite{bader1985atoms}, Voronoi deformation density (VDD) charges \cite{bickelhaupt1996voronoiI, fonseca2004voronoiII}, as well as Hirshfeld (or stockholder) partitioning \cite{hirshfeld1977bonded} and its subsequent variants \cite{bultinck2007hirshfeldI,verstraelen2013hirshfeldE, lillestolen2009hirhfeld-iterative}.
Finally, the last category includes methods based on the fitting of a set of point charges to reproduce an observable quantity, such as the molecular electrostatic potential (ESP). 
However, in these methods, charges often converge to different values upon small changes in the molecular geometry. 
This conformational dependency can be reduced by using multiple conformations during the fitting procedure, with a noticeable improvement in the charge set transferability \cite{reynolds1992atomic}.
Moreover, the charges can converge to significantly different values, especially for buried atoms in the molecule, and nevertheless reproduce the ESP equally well.
For this reason, restraints to regularize the problem are often introduced.
One common strategy to solve this problem is to restrain the charges to zero via a hyperbolic penalty function, as done in the restrained ESP (RESP) approach \cite{bayly1993well}.
Still, the restraint of atomic charges to zero lacks strong physical motivation and comes more from practical considerations, to regularize the problem.

The RESP approach provides an accurate description of the electrostatic properties of molecular systems, which have a direct impact on, e.g., conformational energies \cite{wang2000well, cornell2002application} and intermolecular interactions. 
Therefore, RESP charges are commonly used in the parametrization of classical force fields (FFs) \cite{cornell1995second}, such as in the AMBER FFs \cite{duan2003point, ponder2003force, wang2004development}.
In practice, RESP fitting is performed on small molecules or molecular fragments in the gas phase, using the ESP on grid points around the molecule as a reference.
For example, in the Merz--Singh--Kollman scheme \cite{singh1984merz-singh-koolam_schemeI, besler1990atomicmerz-singh-koolam_schemeII}, the grid points are located on several layers around the molecule, constructed around each atom as a union of spheres of increasing radius, i.e., from 1.4 up to 2.0 times the van der Waals radius.
Empirical observations have shown that combining the Hartree--Fock or B3LYP method with the 6-31G* basis set overestimates the polarization of the system in the gas phase, which makes it often suitable for use in the condensed phase \cite{bayly1993well, cornell1995second}.
However, there are no rigorous guarantees that ensure the transferability of gas-phase derived charge values to a large variety of complex electrostatic environments. 

In the context of quantum mechanical/molecular mechanical (QM/MM) molecular dynamics (MD) simulations,  \citeauthor{laio2002dresp}\cite{laio2002dresp} introduced a scheme where RESP charges are dynamically generated (D-RESP) on the fly during the simulation, using the MM atom position as probe sites for the ESP.
This method can provide a set of atomic point charges able to accurately reproduce the total molecular moments, as well as the ESP generated by the QM subsystem on the MM atoms. 
In this work, we extended the original D-RESP approach \cite{laio2002dresp} to fit atom-centered electric multipole moments of arbitrary order, resulting in the extended D-RESP (xDRESP) approach.
An additional benefit of the presented xDRESP method is its implementation within the MiMiC framework for multiscale modeling \cite{olsen2019mimic, antalik2024mimic}, which combines the capabilities of different external programs responsible for computing energy and force contributions of different subsystems.
MiMiC itself only calculates the subsystem interactions and enables communication between the different external programs.
The modular design of MiMiC provides access to the xDRESP approach using any combination of available QM and MM external programs in QM/MM MD simulations.

In the following, we discuss the theoretical details of the xDRESP approach, as well as practical implications of its implementation within the MiMiC framework (Sec.~\ref{sec:theory}). 
We then focus on different examples to validate the method and show potential applications for different chemical problems (Sec.~\ref{sec:results}), and we conclude by highlighting possible new methodological advantages enabled by the implementation of the xDRESP approach (Sec.~\ref{sec:conclusions}).

\section{\label{sec:theory}Theory and Methods}
\subsection{\label{sec:dresp}Extended D-RESP approach}
The xDRESP atom-centered electric multipole moments can be obtained by solving a least-squares problem on the ESP generated by the QM subsystem on the MM atoms ($V^{\mathrm{QM \to MM}}$) with additional restraints on the values of the point charges ($q^{\mathrm{ref}}$), with the option of imposing additional constraints ($\sigma$).
The loss function for such optimization takes the form
\begin{equation}
\label{eq:leastsquare}
    L = \sum_{p=1}^{N^\mathrm{SR}}{\left(V_p^{\mathrm{xDRESP}} - V_p^{\mathrm{QM \to MM}} \right)}^2 + w_R \sum_{i=1}^{N^\mathrm{QM}}{\left( M_i^{\mathrm{[0]}} - q_i^{\mathrm{ref}}\right)}^2 + \sigma,
\end{equation}
where $N^{\mathrm{SR}}$ is the number of short-range (SR) atoms, i.e., a subset of MM atoms surrounding the QM subsystem used as probe sites for the fit of the ESP, $V_p^{\mathrm{QM \to MM}}$ is the value of the reference ESP on the $p$th MM SR site, i.e., the ESP generated by the QM electrons and nuclei at the position of the $p$th MM atom, $w_R$ is the weighting factor for the charge restraint, $M_i^{\mathrm{[0]}}$ is the $0$th order electric multipole on the $i$th QM atom, i.e., the atomic charge, and $q_i^{\mathrm{ref}}$ is its reference value.
The ESP  on the $p$th site generated by the xDRESP multipoles corresponds to
\begin{equation}
\label{eq:dresp_pot}
    V_p^{\mathrm{xDRESP}} = \sum_{i=1}^{N^{\mathrm{QM}}} \sum_{\left|\alpha\right|=0}^{\Lambda} \frac{\left(-1\right)^{\left|\alpha\right|}}{\alpha!}M_i^{\left[\alpha\right]} T^{\left[\alpha\right]}{\left(\mathbf{R}_p, \mathbf{R}_i\right)},
\end{equation}
where $T^{\left[\alpha\right]}{\left(\mathbf{R}_a, \mathbf{R}_b\right)} = \partial^{\alpha}_{\mathbf{R}_b}{\left|\mathbf{R}_b - \mathbf{R}_a\right|}^{-1} = \partial^{\left|\alpha\right|} / \partial R_{b,x}^{\alpha_x} \partial R_{b,y}^{\alpha_y} \partial R_{b,z}^{\alpha_z} \left|\mathbf{R}_b - \mathbf{R}_a\right|^{-1}$ is the interaction tensor.
The multi-index $\alpha = (\alpha_x\;\alpha_y\;\alpha_z)$ indicates the Cartesian components, and $\left|\alpha\right| = \alpha_x + \alpha_y + \alpha_z$ and $\alpha! = \alpha_x!\cdot\alpha_y!\cdot\alpha_z!$, while $\Lambda$ is the maximum order included in the xDRESP multipole expansion, e.g., for $\Lambda=0$ only $\alpha = (0\;0\;0)$ is included (atomic charges), for $\Lambda=1$, $(0\;0\;0)$, $(1\;0\;0)$, $(0\;1\;0)$, and $(0\;0\;1)$ are included (atomic charges and dipoles), and so on.

The charge restraints help to regularize the problem, similarly to what is done in the RESP approach \cite{bayly1993well}.
However, in the original D-RESP implementation \cite{laio2002dresp}, the penalty function is quadratic and the restraint is to the Hirshfeld charges computed at each MD step \cite{hirshfeld1977bonded}, providing a more physically-motivated reference value than zero, which is commonly used in the RESP approach.
Nevertheless, our implementation remains flexible, allowing the user to specify a fixed set of reference point charges.
The possibility of using charges from different schemes (Mulliken \cite{mulliken1955electronic}, Bader \cite{bader1985atoms}, etc.) as a reference can be easily added, provided that the external QM program supports the desired scheme.

Optionally, additional constraints can be included with the method of Lagrange multipliers.
In the case of constraints on the molecular charge ($Q_{\mathrm{tot}}^{\mathrm{QM}}$) and molecular dipole components ($D^{\mathrm{QM}}_{\mathrm{tot, \xi}}$, $\xi\in\left\{x,y,z\right\}$), such constraints can be expressed as 
\begin{equation}
\label{eq:leastsquare_constraints}
 \sigma = \lambda_Q\left(\sum_{i=1}^{N^\mathrm{QM}} M_i^{\mathrm{[0]}} - Q_{\mathrm{tot}}^{\mathrm{QM}}\right) + \sum_{\xi\in\left\{x,y,z\right\}}  \sum_{\left|\beta\right|=1} \lambda_{D, \xi} \;\beta_\xi \left[\sum_{i=1}^{N^{\mathrm{QM}}}\left( M_i^{\left[0\right]}R_{i, \xi} + M_{i}^{\left[\beta\right]}\right) - D^{\mathrm{QM}}_{\mathrm{tot, \xi}}\right],
\end{equation}
where $\lambda_Q$ and $\lambda_{D, \xi}$ are the Lagrange multipliers corresponding to the constraints on the charge or on the dipole components, which are the only optional constraints currently implemented.
However, any additional constraints, e.g., to higher-order molecular multipole moment components, can be introduced following the same logic.

Minimization with respect to electric multipole moments $M_i^{\left[\alpha\right]}$ and Lagrange multipliers $\lambda_Q$ and $\lambda_{D, \xi}$ (detailed derivation in Appendix~\ref{sec:eq_derivation}), leads to a system of linear equations that can be expressed in matrix form as
\begin{equation}\label{eq:matrix_problem}
\begin{pmatrix}
\mathbf{A}^{\left[0\right]\left[0\right]} & \mathbf{A}^{\left[0\right]\left[1\right]} & \dots & \mathbf{A}^{\left[0\right]\left[\Lambda\right]} & \mathbf{1}& \mathbf{R}_x& \mathbf{R}_y& \mathbf{R}_z\\
\mathbf{A}^{\left[1\right]\left[0\right]} & \mathbf{A}^{\left[1\right]\left[1\right]} & \dots & \mathbf{A}^{\left[1\right]\left[\Lambda\right]}& \mathbf{0}& \mathbf{1}& \mathbf{1}& \mathbf{1}\\
\vdots & \vdots & \ddots & \vdots& \mathbf{0}& \mathbf{0}& \mathbf{0}& \mathbf{0}\\
\mathbf{A}^{\left[\Lambda\right]\left[0\right]} & \mathbf{A}^{\left[\Lambda\right]\left[1\right]} & \dots & \mathbf{A}^{\left[\Lambda\right]\left[\Lambda\right]}& \mathbf{0}& \mathbf{0}& \mathbf{0}& \mathbf{0}\\
\mathbf{1}^\mathrm{T} & \mathbf{0}^\mathrm{T} & \mathbf{0}^\mathrm{T} & \mathbf{0}^\mathrm{T} & 0 & 0 & 0 & 0\\
\mathbf{R}_x^\mathrm{T} & \mathbf{1}^\mathrm{T} & \mathbf{0}^\mathrm{T} & \mathbf{0}^\mathrm{T} & 0 & 0 & 0 & 0\\
\mathbf{R}_y^\mathrm{T} & \mathbf{1}^\mathrm{T} & \mathbf{0}^\mathrm{T} & \mathbf{0}^\mathrm{T} & 0 & 0 & 0 & 0\\
\mathbf{R}_z^\mathrm{T} & \mathbf{1}^\mathrm{T} & \mathbf{0}^\mathrm{T} & \mathbf{0}^\mathrm{T} & 0 & 0 & 0 & 0
\end{pmatrix}
\begin{pmatrix}
\mathbf{M}^{\left[0\right]}\\\mathbf{M}^{\left[1\right]}\\ \vdots\\ \mathbf{M}^{\left[\Lambda\right]}\\\frac{1}{2}\lambda_Q\\\frac{1}{2}\lambda_{D, x}\\\frac{1}{2}\lambda_{D, y}\\\frac{1}{2}\lambda_{D, z} 
\end{pmatrix}
= \begin{pmatrix}
\mathbf{b}^{\left[0\right]}\\\mathbf{b}^{\left[1\right]}\\ \vdots\\ \mathbf{b}^{\left[\Lambda\right]}\\Q_{\mathrm{tot}}^{\mathrm{QM}}\\ D_{\mathrm{tot}, x}^{\mathrm{QM}}\\ D_{\mathrm{tot}, y}^{\mathrm{QM}}\\ D_{\mathrm{tot}, z}^{\mathrm{QM}}
\end{pmatrix},
\end{equation}
where $\mathbf{R}_\xi$ is a vector containing the $\xi$-th Cartesian coordinate components of the QM atoms, and $\mathbf{1}$ and $\mathbf{0}$ represent column vectors of ones and zeros, respectively.
The matrix on the left-hand side, \emph{influence matrix}, is a symmetric square matrix of size $n\times n$, with $n = N^{\mathrm{QM}}\sum_{\alpha=0}^\Lambda 3^\alpha + N_{\mathrm{constraints}}$, and each $\mathbf{A}^{\left[\alpha\right]\left[\beta\right]}$ element is a square matrix of size $N^{\mathrm{QM}}\times N^{\mathrm{QM}}$ with elements
\begin{equation}
A_{i,j}^{\left[\alpha\right]\left[\beta\right]} 
    =  \frac{(-1)^{\left|\alpha\right|+ \left|\beta\right|}}{\alpha!\cdot\beta!} \sum_{p=1}^{N^{\mathrm{SR}}}T^{\left[\alpha\right]}{\left(\mathbf{R}_p, \mathbf{R}_i\right)} T^{\left[\beta\right]}{\left(\mathbf{R}_p, \mathbf{R}_j\right)}  +w_R\delta_{\beta,0}\delta_{i,j}.
\end{equation}
The vector on the right-hand side, \emph{target vector}, contains $\mathbf{b}^{\mathrm{\beta}}$ vectors of size $N^{\mathrm{QM}}\times 1$ with elements
\begin{equation}
b_j^{\left[\beta\right]} =\frac{\left(-1\right)^{\left|\beta\right|}}{\beta!}\sum_{p=1}^{N^{\mathrm{SR}}}V_p^{\mathrm{QM \to MM}}T^{\left[\beta\right]}{\left(\mathbf{R}_p, \mathbf{R}_j\right)} +w_R\delta_{\beta,0} \delta_{i,j}q_j^{\mathrm{ref}}.
\end{equation}

\subsection{\label{sec:mimic-dresp}xDRESP scheme within the MiMiC Framework}

The main idea behind MiMiC \cite{olsen2019mimic, antalik2024mimic} is to provide an interface between external programs that perform specific calculations on individual subsystems at different levels of theory or resolutions, e.g., QM or MM.
MiMiC handles the communication between the external programs and computes the interactions between the subsystems.
Once an external program is interfaced with MiMiC, it can be automatically used in conjunction with other programs, provided that the desired type of subsystem interaction is supported.
In the specific case of QM/MM, MiMiC currently supports CPMD \cite{cpmd_free} and CP2K \cite{Kuhne2020} as QM programs \cite{olsen2019mimic, antalik2024mimic}, as well as GROMACS \cite{abraham2015} and OpenMM \cite{eastman2023openmm} as MM programs \cite{viacheslav2019extreme, levy2025openmm, levy2025chimia}. 
More programs are to be introduced soon, following the ongoing efforts to provide interfaces to Quantum ESPRESSO \cite{Carnimeo2023}, DFT-FE \cite{Das2022}, and TinkerHP \cite{thp2}.
The xDRESP approach presented in this work is immediately available for an arbitrary combination of QM and MM programs. 
For example, some of the simulations presented in this manuscript were performed using GROMACS in combination with CPMD, while for others, we used CP2K.

On the external QM program side, the only modification needed to perform an xDRESP QM/MM calculation is to communicate the reference charges to be used as restraints at each MD step.
In MiMiC, the xDRESP calculation can be controlled by user-specified parameters, such as the maximum order for the atom-centered multipoles, e.g., 0 to fit point charges or 1 to fit atomic charges and dipoles.
It is also possible to specify the restraint weight or the type of charges to be used as restraints, as they can be either Hirshfeld charges computed at each MD step or a fixed set of user-defined point charges.
Notably, in the original approach, the D-RESP charges have been shown to reproduce the total molecular moments, even though no explicit constraint on these moments was introduced \cite{laio2002dresp}, and the same can be expected from the xDRESP scheme.
Additionally, we provide the possibility of specifying optional constraints to explicitly ensure that the xDRESP multipoles exactly reproduce the total multipole moments of the QM subsystem.
At the moment, supported constraints include those for the total QM charge and QM dipole, but the same logic can also be used to constrain higher-order total multipoles.

As shown in Eq.~\ref{eq:dresp_pot}, to solve the least-squares problem, we use as a reference $V^{\mathrm{QM \to MM}}$, i.e., the ESP from the QM charge distribution on a set of SR MM atoms.
MiMiC computes the electrostatic QM/MM interactions using a generalized version of the electrostatic QM/MM coupling scheme by \citeauthor{laio2002hamiltonian}\cite{laio2002hamiltonian}, in which the MM atoms are divided into SR and long-range (LR) domains.
The electrostatic QM/MM interaction is then computed exactly for the SR atoms, while the LR atoms interact with the multipoles of the QM subsystem.
\citeauthor{olsen2019mimic}\cite{olsen2019mimic} generalized the original scheme to use multipole expansions of arbitrary order, further improving the accuracy of the LR contribution and hence enabling the selection of smaller SR regions, with a significant gain in computational efficiency.
Since, during a QM/MM MD simulation, $V^{\mathrm{QM \to MM}}$ acting on SR atoms has to be computed in any case to evaluate the QM/MM forces to propagate the system, we use this set of SR atoms to perform the xDRESP fit, thereby avoiding any overhead from the computation of the reference potential on the probe sites.

\subsection{\label{sec:systems}Computational Details}

\subsubsection{Studied Systems}
\begin{figure}[t]
    \centering
    \includegraphics[width=\linewidth]{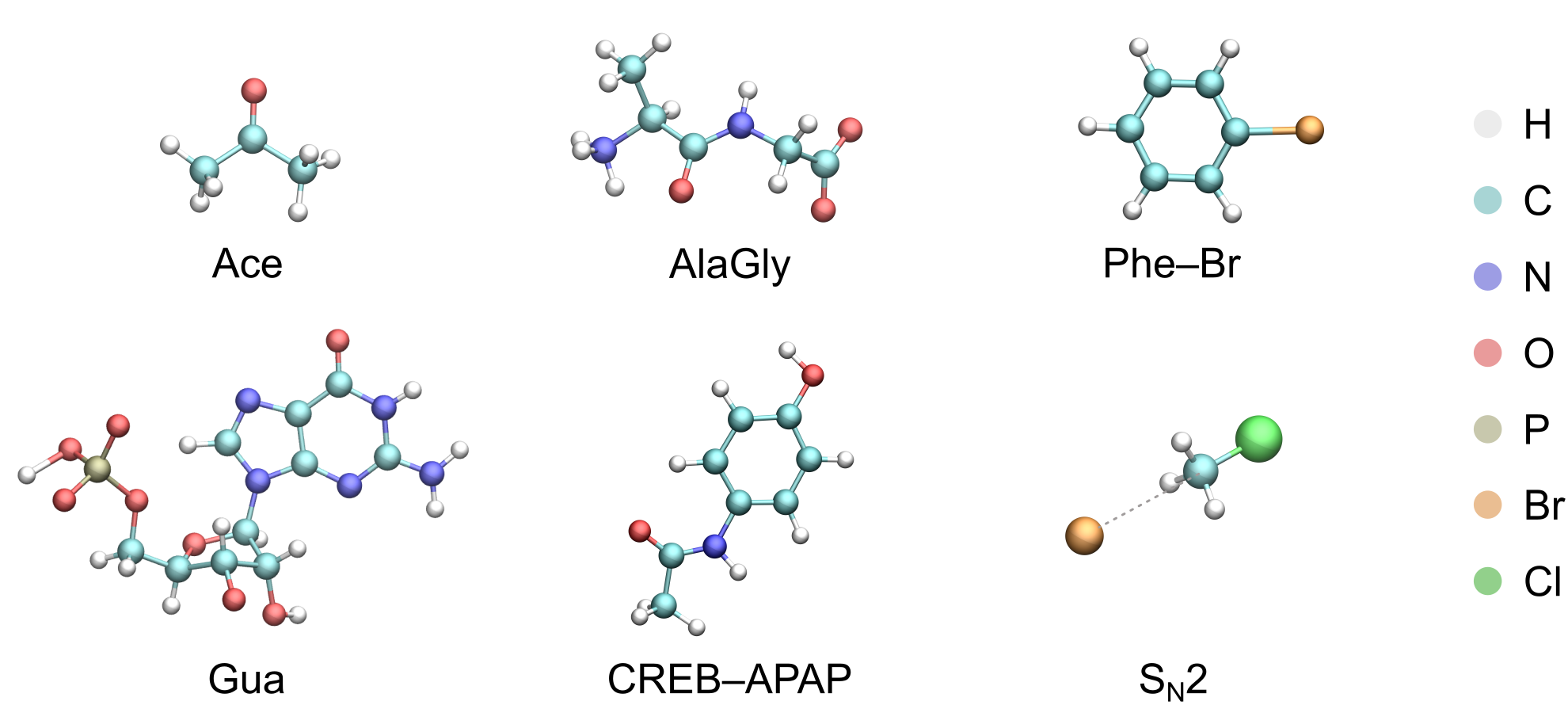}
    \caption{Systems for which the QM/MM MD simulations with xDRESP analysis have been performed.
    Here, only the QM region is represented, but all systems are solvated in water, except for the S$_\mathrm{N}$2 reaction, which has been performed in acetone instead.
    For the CREB--APAP system, the protein target (CREB) is also present in the simulation.}
    \label{fig:systems}
\end{figure}

To test the xDRESP approach in different contexts, we employed a varied range of systems (see Fig.~\ref{fig:systems}).
Details on the MM, QM, and QM/MM parameters for each system are reported in the Supporting Information (SI) in Tab.~\ref{tab:mm_params}--\ref{tab:mimic_params}).

For validation, we used a small system composed of an acetone molecule solvated in water (Ace), for which the equilibrated structure and simulation parameters were taken from our previous work on MiMiC \cite{antalik2025making}.
The acetone molecule was treated with the BLYP functional \cite{Becke1988,Lee1988}, with the parameters for interactions at MM level taken from the OPLS/AA FF \cite{robertson2015OPLSAA}. 
The MM subsystem consists of the surrounding water molecules, which were treated with the TIP3P rigid model \cite{jorgensen1983comparison}.

For the comparison with fixed point-charge FFs, we selected representative systems for different classes of biochemical problems commonly investigated with MD simulations: a dipeptide in water (AlaGly) in its zwitterionic form, enabling us to compare with common protein FFs, a guanine ribose phosphate nucleotide (Gua), for comparison with nucleic acid FFs, and a protein--ligand system (CREB--APAP) corresponding to the bromodomain of the human cAMP response element-binding protein (CREB) in complex with N-(4-hydroxyphenyl)acetamide (APAP), commonly known as paracetamol.
The CREB--APAP system allowed a comparison with general FFs, commonly used for small ligands.
The system was taken from the protein data bank (PDB) X-ray structure ID 4A9K \cite{pdb_4a9k, chung2012fragment} (\SI{1.81}{\angstrom} resolution). All the systems have been parameterized with the LEaP program from AmberTools \cite{case2023ambertools}.
We used the AMBER ff14SB \cite{maier2015ff14sb} with OL15 \cite{galindo2016OL15} and OL3 \cite{zgarbova2011OL3} modifications for DNA and RNA, in combination with the TIP3P rigid water model \cite{jorgensen1983comparison}.
For CREB--APAP, the GAFF \cite{wang2004gaff} has been employed, with point charges generated from a RESP fitting procedure using the HF/6-31G* combination.
After an equilibration at the MM level performed with GROMACS \cite{abraham2015}, we switched to QM/MM with MiMiC using GROMACS \cite{abraham2015} and CP2K \cite{Kuhne2020} as MM and QM external programs, respectively.
For the AlaGly and Gua systems, the QM subsystem corresponded to the biomolecule, while for CREB--APAP, only the APAP ligand was treated at the QM level.
We used the BLYP functional \cite{Becke1988,Lee1988} with density and relative cutoffs determined using standard procedures in CP2K \cite{Kuhne2020}, which resulted in converged values of total energy (Tab.~\ref{tab:qm_params}).
We also converged the QM/MM parameters for the LR/SR electrostatic coupling in MiMiC, i.e., SR cutoff and multipole orders, according to Ref.~\citenum{olsen2019mimic} (Tab.~\ref{tab:mimic_params}). 

Additionally, we studied two systems for which we investigated the effect of including the atomic dipoles in the xDRESP fit.
The first was a fictitious system composed of bromobenzene solvated in water (Ph--Br).
Although bromobenzene is insoluble in water, this solvent was chosen to maximize polarization effects.
We also studied 
the S$_\mathrm{N}$2 reaction \ce{Br^- + CH3Cl -> CH3Br + Cl^-} in acetone solution.
Also for these systems, we followed analogous equilibration procedures, using the GAFF \cite{wang2004gaff} for the solute with AM1-BCC \cite{jakalian2000GAFFAM1BCC, jakalian2002GAFFAM1BCC} point charges.
In case of water solvent, we employed the TIP3P rigid model \cite{jorgensen1983comparison}, while for acetone we used the OPLS/AA FF \cite{robertson2015OPLSAA} (with the same parameters as for the Ace system).
When switching to QM/MM, we used the BLYP functional \cite{Becke1988,Lee1988} and converged QM and QM/MM parameters as for the other systems (Tab.~\ref{tab:qm_params} and \ref{tab:mimic_params}). 

For all QM/MM MD simulations, we used a time step of \num{20}\,$\hbar / E_\mathrm{h}$ ($\sim$ \qty{0.5}{\femto\second}), except for the  S$_\mathrm{N}$2 system where we reduced the time step to  \num{10}\,$\hbar / E_\mathrm{h}$ ($\sim$ \qty{0.25}{\femto\second}).
No bond constraints have been applied to the QM atoms. 
We sampled the NVT ensemble, with two Nosé--Hoover thermostats applied to the QM and MM subsystems, with reference temperatures and coupling frequencies of \SI{300}{\kelvin} and \SI{3000}{\per\cm}, respectively.

\subsubsection{Accuracy metrics}
To evaluate the accuracy of the fitted multipoles, we compared the ESP on the SR atoms generated by the xDRESP multipoles with the one computed at the QM/MM level by MiMiC, i.e., $V^{\mathrm{QM \to MM}}$.
For this purpose, we used two different metrics: the root mean square error (RMSE), 
\begin{equation}
    \mathrm{RMSE} = \sqrt{ \frac{1}{N^{\mathrm{SR}}} \sum_{p=1}^{N^{\mathrm{SR}}} \left( V_{p}^{\mathrm{xDRESP}} - V_{p}^{\mathrm{QM \to MM}} \right)^2 },
\end{equation}
and the mean absolute percentage error (MAPE),
\begin{equation}
    \mathrm{MAPE} = \left.\left( \frac{100}{N^\mathrm{SR}} \sum_{p=1}^{N^{\mathrm{SR}}} \left| \frac{V_{p}^{\mathrm{xDRESP}} - V_{p}^{\mathrm{QM \to MM}}}{V_{p}^{\mathrm{QM \to MM}}} \right|\right)\right|_{V_{p}^{\mathrm{QM \to MM}}>\tau},
\end{equation}
in which we exclude points where the $V_{p}^{\mathrm{QM \to MM}}$ potential is lower than a threshold $\tau$, set to \num{e-3}\,$E_\mathrm{h}/e$ (since the MAPE is inversely proportional to $V_{p}^{\mathrm{QM \to MM}}$).

We also evaluated the accuracy of the reproduced molecular electric multipole moments by comparing the total charge, dipole, and quadrupole with those obtained from the QM subsystem computed by MiMiC from the QM charge density to perform the LR/SR QM/MM coupling.
For the dipole, we report the absolute value of the molecular dipole as a comparison, and for the quadrupole, the six different independent components of its traceless version.

\section{\label{sec:results}Results and Discussion}
To validate the approach, we first illustrate the results of the xDRESP fitting for a simple solute--solvent system, i.e., the Ace system, and then shift our focus to different applications.
Starting by fitting atom-centered point charges and comparing them with commonly used point-charge models in standard FFs, we then move to analyzing the effect of higher-order atom-centered multipoles for systems notoriously hard to model with a simple point-charge model.
Finally, we conclude by demonstrating a potential application of this method, namely, to track ongoing changes in the electronic structure during QM/MM MD simulations.

\subsection{Validation}
At the core of the xDRESP approach lies the least-squares fit to the ESP that the QM subsystem generates on a subset of MM atoms used as probe sites, $V^{\mathrm{QM \to MM}}$.
In MiMiC, the MM atoms are partitioned into SR and LR groups based on the distance from the QM subsystem \cite{olsen2019mimic}, and we used the atoms in the SR group as probe sites for the fit.
In addition to the least-squares problem, Eq.~\ref{eq:dresp_pot} presents a restraint to reference charges, which is introduced to regularize the problem, effectively reducing the fluctuations of the fitted atom-centered multipoles. 

\begin{figure}[!ht]
    \centering
    \includegraphics[width=.5\linewidth]{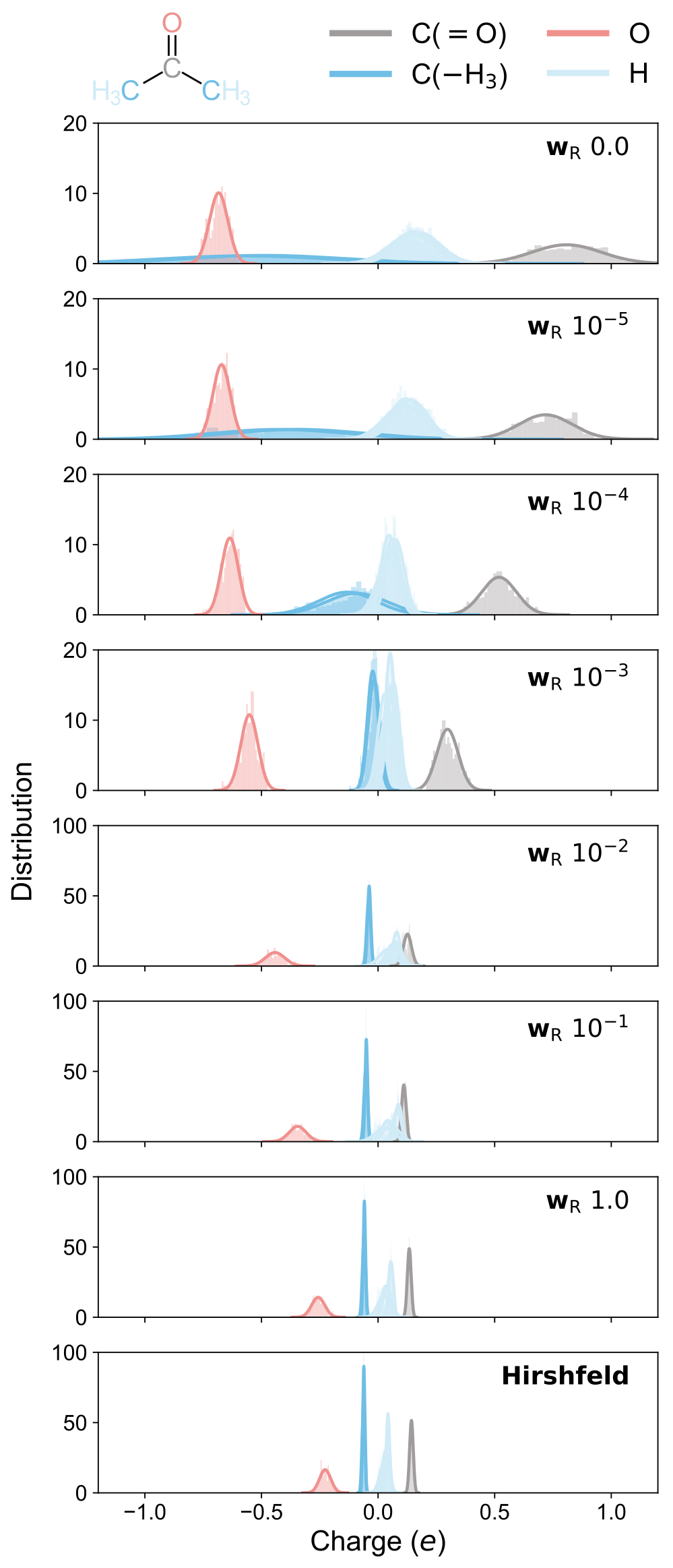}
    \caption{Distribution of xDRESP charges for the Ace sytem during \SI{1}{\pico\second} QM/MM MD with different restraint weights $w_\textrm{R}$ to the Hirshfeld charges.
    In the bottom panel, the reference Hirshfeld charges from the QM external program (CPMD) are reported for comparison.}
    \label{fig:act_dresp_charges}
\end{figure}

We first validated the approach by investigating the effect of charge restraints on the fit of atom-centered charges.
For the Ace system, we systematically increased the restraint weight, $w_\mathrm{R}$, from \num{0.0} to \num{1.0}, i.e., starting with the fit performed without any additional restraint up to imposing restraints of the same weight as the least-squares fit itself.
In Fig.~\ref{fig:act_dresp_charges}, we report the fitted point charges, as well as the Hirshfeld charges used as reference values. 
At small values of the weighting factor ($w_\mathrm{R}$ equal to \num{0.0} and \num{e-5}), the fitted charges present large fluctuations during the dynamics, which are reduced by increasing the weight of the restraint.
Once $w_\mathrm{R}$ is larger than \num{0.1}, the restraint becomes so strong that the charges converge to the reference value, i.e., to the Hirshfeld charges.
A similar trend can be observed when a restraint to a fixed set of point charges is used: in Fig.~\ref{fig:SI_acetone_ref0}, we report the results for an xDRESP fit with $w_\mathrm{R}$ from \num{e-4} to \num{e-1} using zero as a reference charge for all atoms. 
Although for the Ace system the results are similar either with restraints to Hirshfeld charges or to zero, Hirshfeld charges remain preferable, considering they are computed at each MD step directly from the charge density.
Unlike a fixed point-charge model, this approach can capture changes in the electronic density, e.g., during a chemical reaction.

This set of simulations for the Ace system was performed using CPMD as QM external program, for which we implemented the communication of the reference Hirshfeld charges to MiMiC.
We also performed a simulation for the same system, using CP2K as an external QM program, and compared the Hirshfeld charges computed by the two programs  (Fig.~\ref{fig:SI_hirshfeld}).
As expected, with both QM programs, we get similar Hirshfeld charges that are then communicated to MiMiC to perform the xDRESP fitting.
We note that in the case of CP2K, we set the option \texttt{SHAPE\_FUNCTION} to \texttt{DENSITY} to have physically sound values.
This keyword influences the type of shape function used for Hirshfeld partitioning, using in particular multiple Gaussians to expand the charge density of isolated atoms.

\begin{figure}[t]
    \centering
    \includegraphics[width=.6\linewidth]{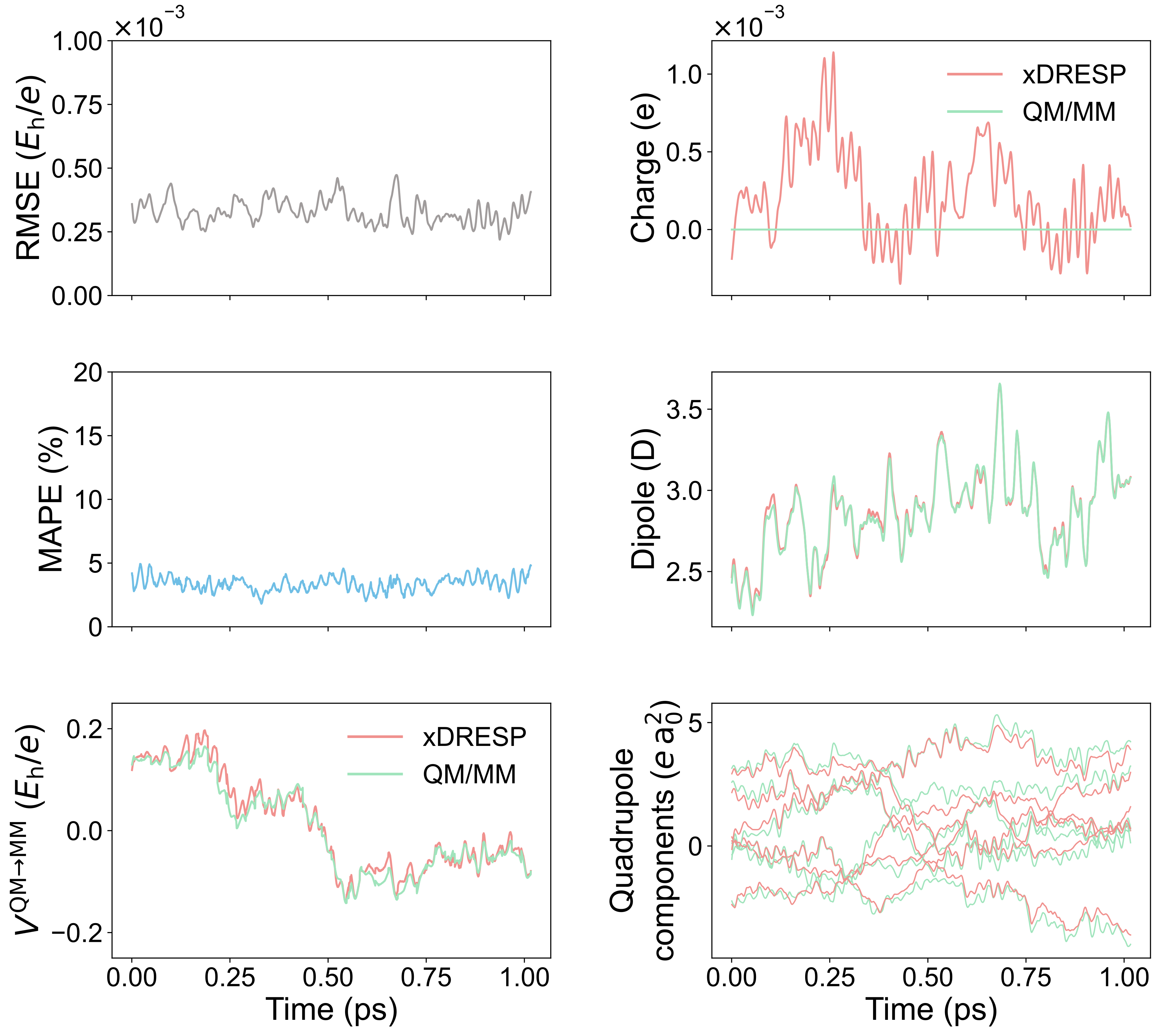}
    \caption{Different metrics used to assess the accuracy of the xDRESP point charge set obtained with $w_\mathrm{R}$ = \num{e-3} for the Ace system.
    The total potential, $V^{\mathrm{QM \to MM}}$, and molecular multipoles are computed during the dynamics for the electrostatic QM/MM coupling, and have been used as reference.}
    \label{fig:act_dresp_metrics}
\end{figure}

Considering intermediate values between $w_\mathrm{R}$ = \num{e-5} and $w_\mathrm{R}$ = \num{0.1}, the fitted charges are stable, presenting physically reasonable average values and fluctuations during the dynamics (average values reported in Tab.~\ref{tab:act_dresp}).
In particular, the oxygen atom bears a negative charge of about \num{-0.5}\,$e$, the carbon and hydrogen atoms of the two methyl groups have charges close to zero, and the carbonyl carbon is slightly positive.
Notably, chemically equivalent atoms in the two methyl groups present similar xDRESP charges.
To quantify the ability to reproduce the ESP, we compared the total $V^{\mathrm{QM \to MM}}$ on the SR atoms computed at each step of the QM/MM MD simulation with that reproduced by the fitted point charges.
 We report in Fig.~\ref{fig:act_dresp_metrics} the resulting metrics for $w_\mathrm{R}$ = \num{e-3}. 
 The same plots for $w_\mathrm{R}$ = \num{e-4} and  \num{e-2} are reported in Fig.~\ref{fig:SI_act_dresp_metrics_pot}.
 For $w_\mathrm{R}$ between \num{e-4} and \num{e-2}, the RMSE is between \num{2e-4}\,$E_\mathrm{h}/e$ and \num{6e-4}\,$E_\mathrm{h}/e$, while the MAPE is between \SI{2.5}{\percent} and \SI{5}{\percent}.
Even if fitted to reproduce the potential, a good set of point charges should ideally also reproduce the molecular multipoles of the system.
We compared the lowest-order molecular multipoles from the QM charge density with the ones generated from the set of xDRESP charges (Fig.~\ref{fig:act_dresp_metrics} for $w_\mathrm{R}$ = \num{e-3}, Fig.~\ref{fig:SI_act_dresp_metrics_pot} for the other $w_\mathrm{R}$ values).
Overall, charge sets generated with the three $w_\mathrm{R}$  values considered well reproduce the total charge, dipole, and quadrupole, especially for $w_\mathrm{R}$ lower than \num{e-3}.

Without imposing explicit constraints, the xDRESP procedure reproduces the total molecular multipole moments well.
However, in some cases, it might be desirable to ensure that the total molecular multipole moments are reproduced exactly.
For this reason, we included the option to impose constraints during the fit.
For the moment, this is limited to the total charge and dipole, with the possibility to extend it to higher molecular multipoles in the future, if needed.
In the SI, we report the results for $w_\mathrm{R}$ = \num{e-3}, where we applied constraints on the total charge of the acetone molecule, or on its total charge and dipole components.
The effect on the xDRESP charges is minimal (Fig.~\ref{fig:SI_constr_dresp}), and the constraint on the total charge seems to have no significant impact on any of the metrics used to evaluate the accuracy of the fitted charges, with the obvious exception of the total charge, which is now reproduced exactly due to the constraint.
In case of additional constraints on the dipole components, we observed a slight decrease in the accuracy to reproduce $V^{\mathrm{QM \to MM}}$ (MAPE increase of $\sim$\SI{2}{\percent}).
We remark that for the Ace system, the inclusion of the additional constraints has minimal effects since the molecular multipoles were already well reproduced, but it illustrates the use of this option in the fitting procedure.

Overall, these results for the Ace system show that the implemented xDRESP approach provides a set of stable point charges that is not only able to reproduce with high accuracy the ESP, $V^{\mathrm{QM \to MM}}$, to which it is fitted, but also the molecular electric multipoles.
The choice of the optimal restraint value may be influenced by the system under study, as well as the number of SR atoms introduced.
For a new system, it might be a good practice to perform short test QM/MM MD simulations with different restraint weights.
Nevertheless, at least in the case of Ace, a value of $w_\mathrm{R}$ = \num{e-3} seems to be a good compromise between stability and ability to reproduce both the ESP and the molecular multipole moments.

\subsection{Comparison with point charge models of classical force fields}

Point charges are fundamental elements of classical FFs, as their values determine the electrostatic interactions.
These interactions are often dominant in many biomolecular problems, so it is of utmost importance to describe them accurately to reproduce the ESP of such systems.
Therefore, we compared the ability of the xDRESP point charges with fixed point-charge models commonly used in standard FFs to reproduce the ESP and molecular multipoles. 
We also tested the accuracy of $\langle\text{xDRESP}\rangle$, a fixed point-charge model obtained by averaging the instantaneous xDRESP charges of each atom during the MD.
This comparison can be used to estimate if a charge set derived for the specific problem is better suited to describe the electrostatic interactions of the system than more transferable charge models used in standard FFs.
Moreover, the comparison between xDRESP and $\langle\text{xDRESP}\rangle$ can be used as an indication of the degree of polarization of the system, which can clarify whether a fixed point-charge model is sufficient.
The $\langle\text{xDRESP}\rangle$ model also represents an initial step towards more comprehensive force-matching approaches to generate an optimal FF specifically fitted for the system under study, such as those proposed in Refs.~\citenum{Maurer2007_FM, Doemer2014_FM, vona2025force_FM}.
In force matching, the electrostatic parameters of a classical FF are fitted using potentials and forces computed during QM or QM/MM MD simulations. 
The fit is performed on a subset of configurations from the MD, and different possibilities exist for obtaining a representative set. 
However, this goes beyond the scope of the current work, and in the results presented in this section, we used $\langle\text{xDRESP}\rangle$ as a first estimate of the potential advantages of using a fixed point-charge model derived from the xDRESP procedure.
Additionally, when parametrizing a FF, whether using force matching or a different approach, it is necessary to assign the same charge to chemically equivalent atoms to define atom types, requiring to impose additional constraints during the charge-fitting procedure.
Such parameters are not yet implemented in the current xDRESP scheme, but can be easily introduced by modifying the least squares problem in Eq.~\ref{eq:leastsquare}.

\begin{figure}[!ht]
    \centering
    \includegraphics[width=.57\linewidth]{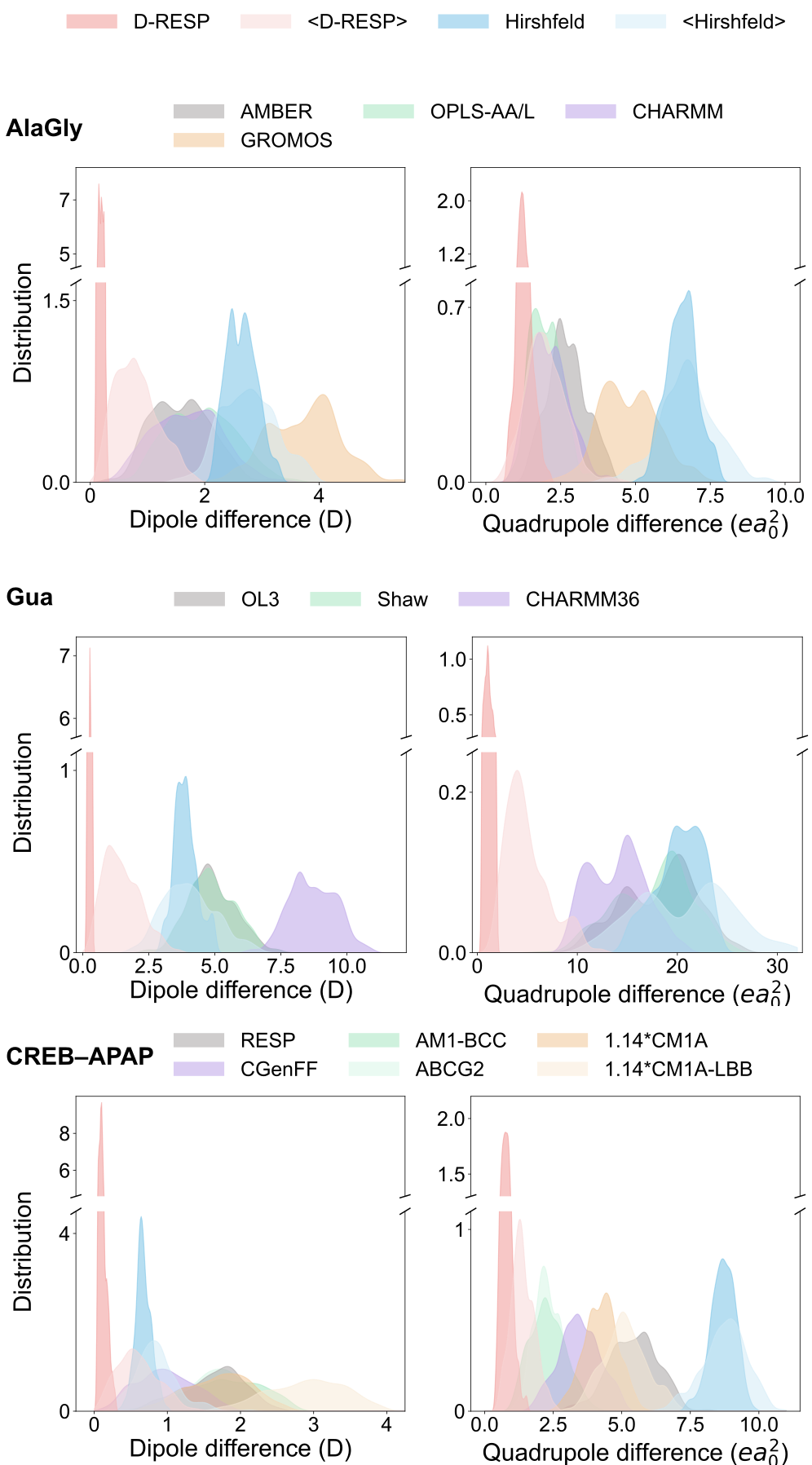}
    \caption{Comparison of the resulting molecular multipole moments (dipole and quadrupole)  from xDRESP charges and from different atomic charge schemes and point-charge models in commonly used classical FFs. 
    Averaged xDRESP charges ($\langle\text{xDRESP}\rangle$) and Hirshfeld charges (instantaneous and averaged) are also shown.
    For three different systems, AlaGly, Gua, and CREB--APAP, we report the data from a \SI{1}{\pico\second} QM/MM MD as a distribution of the difference in the molecular dipole/quadrupole calculated from the point charge model with respect to the one calculated from the QM charge distribution.
    The values are calculated as the Euclidean norm for the difference in the dipole and the Frobenius norm for the difference in the quadrupole. 
    Note that the ranges of the axes change among the different plots.
    }
    \label{fig:point_charge_comparison}
\end{figure}

For all systems, we performed a short \SI{1}{\pico\second} QM/MM MD simulation, along which we fitted the xDRESP point charges on the fly at each time step.
From these, we computed the average xDRESP charges, $\langle\text{xDRESP}\rangle$, and assessed their accuracy in reproducing $V^{\mathrm{QM \to MM}}$ on the SR atoms as well as the overall dipole and quadrupole moments of the QM subsystem.
We compared these results with the ones obtained with both instantaneous and averaged Hirshfeld charges (the latter referred to as $\langle\text{Hirshfeld}\rangle$), as well as with common point-charge models used in standard FFs for different kinds of systems.
The main results from this comparison are summarized in Tab.~\ref{tab:point_charge_comparison} and Fig.~\ref{fig:point_charge_comparison}, while the complete results are reported in the SI (Figs.~\ref{fig:SI_alagly_dresp_charges}--\ref{fig:SI_crepapap_metric_multipoles} and Tab.~\ref{tab:SI_alagly_point_charges}--\ref{tab:SI_apap_point_charges}).
First, we used the AlaGly dipeptide system to compare with common FFs for proteins: the AMBER FF (with the point-charge set shared by FF99SB \cite{hornak2006amberff99SB}, FF14SB \cite{maier2015ff14sb}, and FF19SB \cite{tian2019amberff19SB}), CHARMM (CHARMM27, i.e., CHARMM22 \cite{mackerell1998CHARMM27-CHARMM22} with CMAP for proteins \cite{mackerell2004CHARMM27-CMAP}, as well as CHARMM36 \cite{best2012CHARMM36-proteins}), OPLS-AA/L \cite{kaminski2001OPLSAAL}, and the united atom FF GROMOS96 54A7 \cite{schmid2011GROMOS54a7}.
In the case of the Gua nucleotide system, we employed point-charge sets of  FFs for nucleic acids, which include AMBER FFs OL3 by \citeauthor{zgarbova2011OL3}\cite{zgarbova2011OL3} and its modified version by \citeauthor{tan2018ShawRNA}\cite{tan2018ShawRNA} (Shaw FF), as well as the CHARMM36 FF for RNA \cite{denning2011CHARMM36_RNA}.
Finally, the last system is a protein--ligand (CREB--APAP) complex, where the ligand is described at the QM level, for which we compare with point-charge sets employed in common general FFs specifically designed to describe small molecules. 
Since these are often derived from QM-specific atomic charge schemes, we compare RESP charges from a HF/6-31G* calculation, AM1-BCC \cite{jakalian2000GAFFAM1BCC, jakalian2002GAFFAM1BCC}, and ABCG2 \cite{he2020GAFFABCG2, sun2023GAFFABCG2}, used in the GAFF and GAFF2 general AMBER FFs \cite{wang2004gaff}.
The RESP charges from HF/6-31G* are compatible with both GAFF and GAFF2, but the more efficient AM1-BCC and ABCG2 methods are preferred for large-scale calculations, with GAFF/AM1-BCC and GAFF2/ABCG2 as recommended combinations \cite{amber2025, case2023ambertools}.
The RESP and AM1-BCC charges considered for this study have been calculated as suggested in the Amber 2025 Reference Manual \cite{amber2025, case2023ambertools}, using Gaussian16 \cite{gaussian16_ra03} for the QM reference calculation for the RESP fit, while the ABCG2 charges have been taken from the AMBER small molecule database for GAFF2 \cite{amber2025}.
We also considered the empirical bond-charge increment scheme used in the CHARMM general FF (CGenFF) \cite{vanommeslaeghe2012CHARMMCGENFF_charges1, vanommeslaeghe2012CHARMMCGENFF_charges2}, generated from the web server CHARMM-GUI \cite{jo2008CHARMM-GUI}, as well as the 1.14*CM1A \cite{udier2004OPLS_charges1} and 1.14*CM1A-LBCC \cite{dodda2017OPLS_charges2} used in the OPLS-AA FF \cite{robertson2015OPLSAA}, generated from the web server LigParGen \cite{dodda2017LigParGen} (using the maximum number of optimization iterations, i.e., three).

\begin{table}[!ht]
\caption{Accuracy of the reproduced potential, $V^{\mathrm{QM \to MM}}$, for the AlaGly, Gua, and CREB--APAP systems. 
For the two metrics considered, RMSE and MAPE, the mean value during a \SI{1}{\pico\second} QM/MM MD simulation is reported.
The associated error corresponds to the standard deviation.}
\label{tab:point_charge_comparison}
\centering
\resizebox{0.5\textwidth}{!}{%
\setlength{\tabcolsep}{4pt} 
\renewcommand{\arraystretch}{1.1} 
\begin{tabular}{lcc}
\multicolumn{3}{l}{\textbf{AlaGly}} \\
\hline
Model & RMSE ($\times 10^{-4}\,E_\mathrm{h}/e$) & MAPE (\%) \\ 
\hline
xDRESP & 5.1 $\pm$ 0.5 & 1.8 $\pm$ 0.2 \\
$\langle\text{xDRESP}\rangle$ & 9.3 $\pm$ 2.3 & 5.6 $\pm$ 2.3 \\
Hirshfeld & 25.3 $\pm$ 2.0 & 14.9 $\pm$ 1.4 \\
$\langle\text{Hirshfeld}\rangle$ & 26.5 $\pm$ 3.7 & 15.5 $\pm$ 3.1 \\
AMBER & 15.4 $\pm$ 3.9 & 11.6 $\pm$ 3.7 \\
CHARMM & 16.3 $\pm$ 4.4 & 12.4 $\pm$ 4.1 \\
OPLS-AA/L & 17.9 $\pm$ 4.7 & 14.3 $\pm$ 4.1 \\
GROMOS & 31.6 $\pm$ 4.8 & 25.6 $\pm$ 4.3 \\
\hline
\\
\multicolumn{3}{l}{\textbf{Gua}} \\
\hline
Model & RMSE ($\times 10^{-4}\,E_\mathrm{h}/e$) & MAPE (\%) \\ 
\hline
xDRESP & 4.3 $\pm$ 0.4 & 0.40 $\pm$ 0.05 \\
$\langle\text{xDRESP}\rangle$ & 8.6 $\pm$ 1.9 & 1.3 $\pm$ 0.5 \\
Hirshfeld & 24.9 $\pm$ 0.8 & 4.0 $\pm$ 0.3 \\
$\langle\text{Hirshfeld}\rangle$ & 25.8 $\pm$ 2.9 & 4.1 $\pm$ 0.8 \\
OL3 & 23.4 $\pm$ 3.6 & 4.3 $\pm$ 0.8 \\
Shaw & 23.2 $\pm$ 3.6 & 4.2 $\pm$ 0.7 \\
CHARMM36 & 36.7 $\pm$ 3.7 & 6.7 $\pm$ 0.7 \\
\hline
\\
\multicolumn{3}{l}{\textbf{CREB--APAP}} \\
\hline
Model & RMSE ($\times 10^{-4}\,E_\mathrm{h}/e$) & MAPE (\%) \\ 
\hline
xDRESP & 3.0 $\pm$ 0.4 & 4.7 $\pm$ 0.9 \\
$\langle\text{xDRESP}\rangle$ & 5.2 $\pm$ 1.2 & 13.4 $\pm$ 4.9 \\
Hirshfeld & 14.9 $\pm$ 0.9 & 29.5 $\pm$ 1.4 \\
$\langle\text{Hirshfeld}\rangle$ & 15.4 $\pm$ 1.2 & 31.1 $\pm$ 3.0 \\
RESP & 12.8 $\pm$ 2.0 & 34.8 $\pm$ 6.7 \\
AM1-BCC & 11.3 $\pm$ 2.6 & 33.3 $\pm$ 9.5 \\
ABCG2 & 11.6 $\pm$ 2.1 & 34.9 $\pm$ 7.1 \\
CGenFF & 8.1 $\pm$ 1.6 & 20.9 $\pm$ 5.6 \\
1.14*CM1A & 12.7 $\pm$ 2.0 & 37.3 $\pm$ 7.7 \\
1.14*CM1A-LBB & 17.6 $\pm$ 2.4 & 51.2 $\pm$ 9.6 \\
\hline
\end{tabular}
}
\end{table}

Starting with the ability to reproduce $V^{\mathrm{QM \to MM}}$, the accuracy varies across different point charge sets and systems (Tab.~\ref{tab:point_charge_comparison}).
As expected, the instantaneously fitted xDRESP charges show the lowest RMSE and MAPE among all charge sets, as they have been directly fitted to reproduce the QM potential on these sites at every MD step.
Moving to the fixed $\langle\text{xDRESP}\rangle$ point-charge set, it results in only a small decrease in accuracy, with roughly double the RMSE for all systems, yet still well below the errors of other charge representations.
This corresponds to an increase in the MAPE of only a few percent for AlaGly and Gua, while the effect is larger for the CREB--APAP case.
For this system, the noticeable difference between the performance of the instantaneous xDRESP charges and $\langle\text{xDRESP}\rangle$ (increase in MAPE from $\sim$\SI{5}{\percent} to $\sim$\SI{13}{\percent}) points to the relative importance of polarization effects that can limit transferability of point charge models.
This is particularly challenging for generalized FFs that need to describe ligands across different regions of chemical space.
The Hirshfeld charges, which were used as a reference for the xDRESP fitting, perform significantly worse than xDRESP for both instantaneous and averaged $\langle\text{Hirshfeld}\rangle$ sets.
The majority of FF point-charge sets tested reproduce the potential with roughly the same or lower error than Hirshfeld charges.
In the case of AlaGly, AMBER, CHARMM, and OPLS-AA/L reproduce the potential within about a \SI{10}{\percent} MAPE, and, interestingly, also GROMOS FF, which is a united atoms FF, provides a rather good description of the ESP (MAPE $\sim$\SI{25}{\percent}).
For the Gua system, FF point-charges provide a similar accuracy as Hirshfeld with absolute errors of the same magnitude as for the AlaGly dipeptide and with even more reduced relative percentage errors of roughly 4--\SI{6}{\percent}, possibly due to the fact that the ESP has larger absolute values.
Finally, for CREB--APAP, all point-charge sets present similar RMSE (as for the other two test systems), but all exhibit larger MAPE compared to the previous systems, with CGenFF being noticeably better than other FFs. 
In contrast, the 1.14CM1A-LBB model exhibits significantly higher MAPE, almost double that of the other sets.

Regarding the ability to reproduce molecular multipole moments, most point-charge sets used in FFs reproduce the total charge by construction.
This is not the case for the xDRESP and $\langle\text{xDRESP}\rangle$ sets since no explicit constraint is imposed, but they still result in a total charge within \num{2e-3}\,$e$ from the QM reference.
Concerning the total dipole and quadrupole moments, the point charges from the instantaneous xDRESP fit provide the best estimate for the QM values (Fig.~\ref{fig:point_charge_comparison}).
Even though the FF point-charges manage to reproduce the fluctuations in these quantities for the AlaGly system in a qualitative manner, they struggle to do so for the other two test systems.
Particularly for CREB--APAP, it is important to note that the absolute value of the dipole is much smaller compared to the other two systems.
This means that even if the absolute errors seem similar or lower than those for AlaGly, they are still significant.
In the case of the Gua system, the mean difference in the dipole is about \num{4}\,D, corresponding to an error of $\sim$\SI{13}{\percent} in the total dipole.
In contrast to these observations, the xDRESP-derived charges perform reasonably well even in their averaged form.

Overall, the results of this section, even though limited to a few representative examples, show that the xDRESP representation and partially also the $\langle\text{xDRESP}\rangle$ fixed point-charge model can accurately reproduce $V^{\mathrm{QM \to MM}}$ as well as the molecular multipoles.
The fixed point-charge models of commonly used biomolecular FFs for proteins and nucleic acids also show good agreement with the QM data, especially in reproducing the potential.
Yet, we observed larger discrepancies in the case of point-charge models used in generalized FFs. 
This is not particularly surprising, given that such models are designed for broad applicability across different regions of chemical space.
Still, given the crucial role of electrostatic interactions, e.g., in ligand--protein complexes, it is essential to validate the point-charge model used for a specific system to ensure accuracy and confidence in the model's quantitative predictive power.

\subsection{Effect of higher-order multipoles: halogenated benzenes}
Halogens represent a group of elements that are of particular interest in rational drug design, due to their capability to help in optimization and refinement of ligand-receptor interactions \cite{mendez2017looking, ali2025highlights}.
However, these compounds pose severe challenges for computational modeling due to the anisotropy of the charge distribution of halogen atoms, resulting in the $\sigma$-hole, i.e., a positively charged electrostatic region extending from a carbon--halogen bond leading to so-called halogen bonding effects.~\cite{politzer2013halogen}
Since classical FFs based on fixed atomic point charges are not able to accurately capture the electrostatics of $\sigma$-hole systems, it is common to use different strategies, such as the inclusion of a charged virtual site on the halogen, as, e.g., in the CGenFF FF \cite{gutierrez2016parametrization}, or the use of polarizable FFs \cite{ns2015polarizable, lin2018polarizable}.

\begin{figure}[t]
    \centering
    \includegraphics[width=\linewidth]{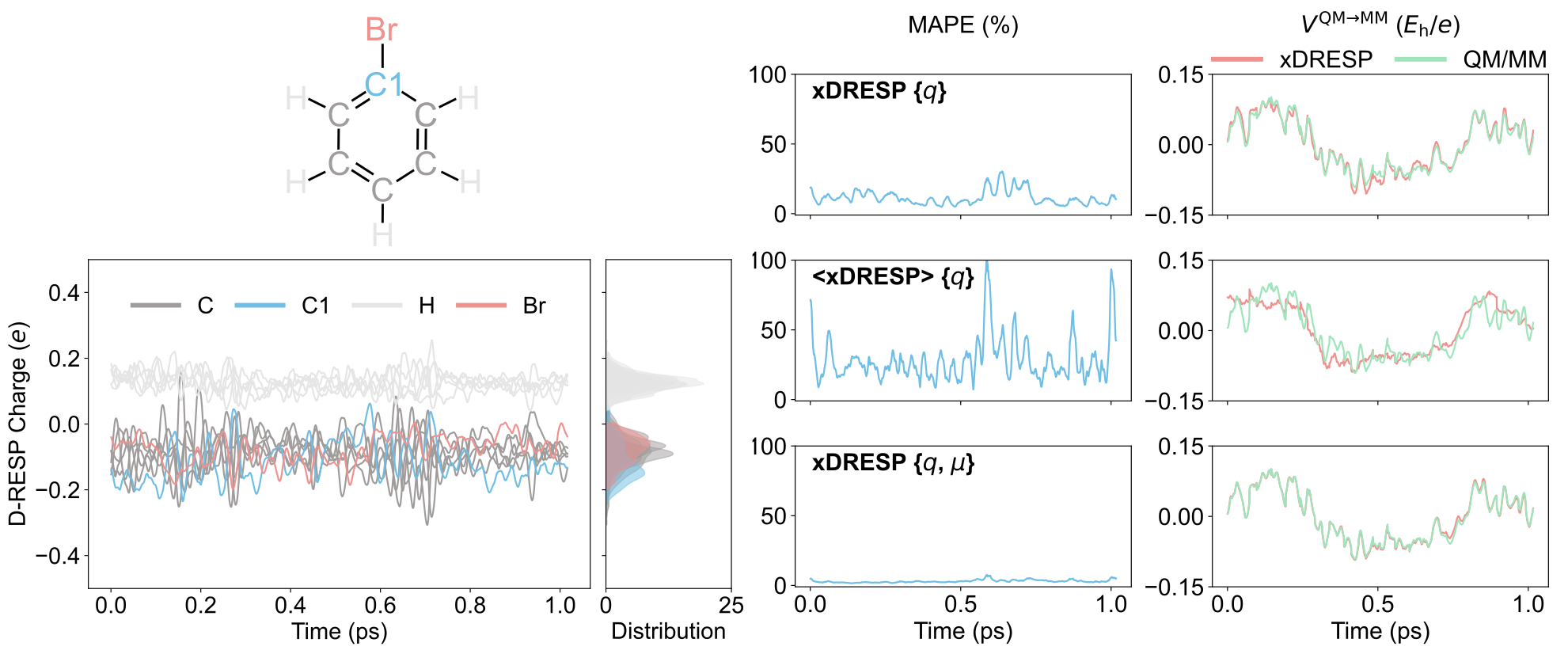}
    \caption{xDRESP charges for the Phe--Br molecule during \SI{1}{\pico\second} QM/MM MD with a restraint to the Hirshfeld charges $w_\textrm{R}$=\num{e-3}, fitting atomic charges (left) and comparison of the ability in reproducing the potential, $V^{\mathrm{QM \to MM}}$, with xDRESP, the fixed point-charge variant $\langle\text{xDRESP}\rangle$ and a simultaneous xDRESP fit of atomic charges and dipoles xDRESP\{$q$, $\mu$\} (right).}
    \label{fig:PhBr_results}
\end{figure}

In the following, we investigate the effects of the inclusion of permanent multipole moments in the case of bromobenzene (Ph--Br), as a representative member of the class of halobenzenes.
In particular, we monitored the behavior of the fitted xDRESP charges and their impact on reproducing $V^{\mathrm{QM \to MM}}$ during a \SI{1}{\pico\second} QM/MM MD simulation of Ph--Br solvated in water.
As shown in Fig.~\ref{fig:PhBr_results}, we observed large fluctuations of the charges even if using the same restraint weight as for the other test system ($w_\textrm{R}$=\num{e-3}), especially for the C and Br atoms, which are to be expected considering the anisotropy of the charge distribution of the system.
This also demonstrates how the xDRESP fit can serve as an intuitive on-the-fly diagnostic tool for evaluating the degree to which a system is polarized, suggesting a potential insufficiency of a fixed point-charge model.
On top of that, we evaluated the accuracy of the fixed $\langle\text{xDRESP}\rangle$ point-charge set and analysed the impact of also fitting the first-order atom-centered multipoles, i.e., atomic dipoles.

Unlike in all previously considered systems, even the instantaneously fitted xDRESP point charges present larger errors in reproducing $V^{\mathrm{QM \to MM}}$ as shown by the average  MAPE of \SI{11.5}{\percent}, with its instantaneous value peaking up to \SI{30}{\percent}.
In comparison, the MAPE for xDRESP point charges ranged between \SI{0.4}{\percent} and \SI{4.7}{\percent} for the other systems. 
The accuracy further decreases when we consider the fixed point-charge set, $\langle\text{xDRESP}\rangle$, for which the average MAPE value is \SI{28}{\percent}, with peaks reaching \SI{100}{\percent}, corresponding to an absolute  RMSE of \num{1.9e-3}\,$E_\mathrm{h}/e$.
However, the accuracy in reproducing the potential improves dramatically, once we fit both xDRESP charges and dipole moments, with an average MAPE of \SI{3}{\percent}, which translates to an improvement of one order of magnitude in terms of RMSE.

When fitting xDRESP charges only, the molecular multipoles are well-reproduced by both the instantaneous xDRESP point charges and the fixed $\langle\text{xDRESP}\rangle$ variant  (Fig.~\ref{fig:SI_Brbnz_metric_multipoles_complete}). 
However, this is no longer the case when simultaneously fitting atomic point charges and dipoles, as the resulting set of atom-centered multipoles reproduces the molecular dipole and quadrupole less accurately.
This problem can be partially overcome by adding a constraint on the total charge and dipole, which, by construction, ensures that the fitted atomic charges and dipoles reproduce the molecular ones.
Despite this improvement, the resulting multipoles present a larger MAPE than the ones from an unconstrained fit, with the molecular quadrupole reproduced rather poorly, exhibiting large deviations from the QM reference.

The inclusion of the higher-order atom-centered multipoles enables a more accurate reproduction of the ESP by providing additional degrees of freedom for the xDRESP fit. 
However, unlike in the point-charge case, in which a restraint is applied to the Hirshfeld charges, we do not introduce any additional restraints for higher-order multipoles.
This results in large fluctuations in the atomic dipoles, especially for the aromatic \ce{C} atoms (Fig.~\ref{fig:SI_dresp_Brbnz_dipole_fit_dipole}).
These observations suggest that further restraints, similar to the ones imposed on the atomic charges, might be necessary to regularize the problem. 
For example, the multipoles could be restricted to the atom-centered multipoles estimated from the Hirshfeld atomic multipoles method \cite{hirshfeld1977bonded}, or from Stone's distributed multipole analysis \cite{stone1981STONE1, stone1985STONE2, stone2013STONE3}.
What makes this potentially less straightforward to implement is that, unlike point-charge models, atom-centered multipoles of higher order are not commonly available in most QM programs. 
A possible way to overcome this is to compute reference multipoles directly in MiMiC, thus removing any dependence on the external programs for the reference atomic multipole moments.
Nonetheless, this goes beyond the current implementation and will be subject to future investigation.

\subsection{On-the-fly tracking of changes in the charge distribution:\\
\texorpdfstring{S$_\mathrm{N}$2}{SN2} reaction}
Atom-centered multipoles can serve as a useful on-the-fly analysis tool for molecular systems, able to simplify the detection and interpretation of complex electronic rearrangements \cite{langner2016tracking}.
To demonstrate this, we performed a study using the thermodynamic integration (TI) technique \cite{carter1989constrained, sprik1998free} to investigate a prototypical S$_\mathrm{N}$2 bimolecular nucleophilic substitution, a fundamental reaction in textbook organic chemistry, which is often encountered in complex biochemical reaction mechanisms \cite{hase1994simulations, xie2016rethinking}.
The reaction rates of S$_\mathrm{N}$2 reactions are strongly affected by the nature of the solvent, which tends to increase the activation energy with respect to the gas phase \cite{valverde2022free}.
For our investigation, we chose a typical example of this reaction, i.e., \ce{Br^- + CH3Cl -> CH3Br + Cl^-} taking place in acetone, with the reaction participants treated at the QM level, while the solvent is modeled with the OPLS/AA FF \cite{robertson2015OPLSAA}.

The calculation of free-energy profiles with the TI technique is based on a series of constrained MD simulations.
In the case of distance constraints, the free energy can be estimated as the integral of the ensemble average of the constraint force along the constrained reaction coordinate defined by a suitably chosen collective variable (CV) \cite{carter1989constrained, sprik1998free}.
For this study, we selected the distance difference between the \ce{C}--\ce{Cl} and \ce{C}--\ce{Br} distances, i.e., $\mathrm{CV}=d(\ce{C-Cl}) - d(\ce{C-Br})$.

\begin{figure}[!ht]
    \centering
    \includegraphics[width=0.5\linewidth]{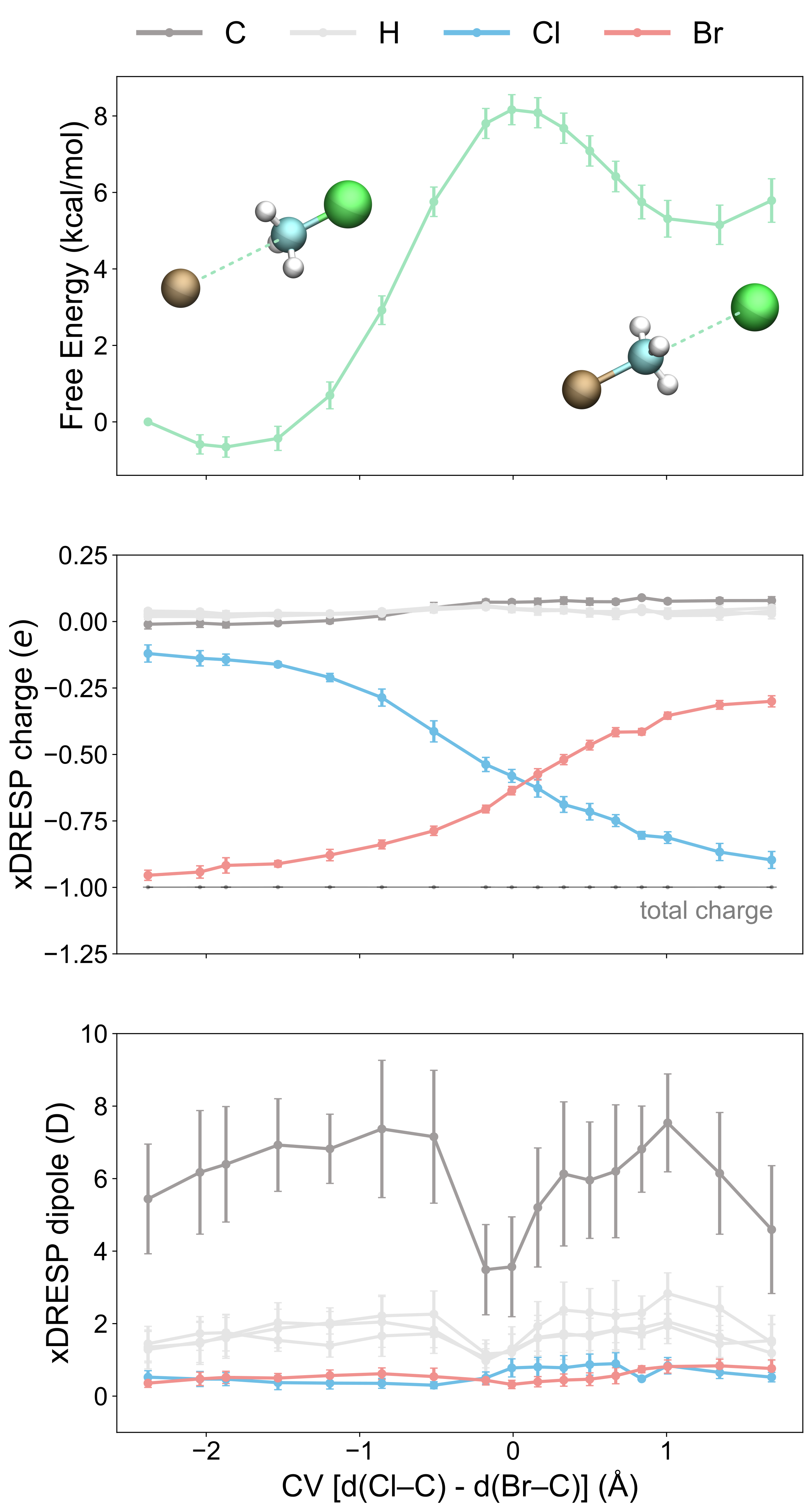}
    \caption{Changes along the S$_\mathrm{N}$2 (\ce{Br^- + CH3Cl -> CH3Br + Cl^-}) reaction. Free energy profile obtained via thermodynamics integration (top panel); values of the fitted xDRESP atomic charges (middle panel) and atomic dipoles (bottom panel) along the reaction pathway.}
    \label{fig:sn2_results}
\end{figure}

The resulting free-energy profile associated with the reaction is reported in Fig.~\ref{fig:sn2_results}: the reaction presents two minima, corresponding to the formation of the \ce{ClCH3} and \ce{BrCH3} molecules, separated by an energy barrier of $\sim$9\,kcal$\,\mathrm{mol}^{-1}$ for the reaction \ce{Br^- + CH3Cl -> CH3Br + Cl^-}, and $\sim$4\,kcal$\,\mathrm{mol}^{-1}$ in the opposite direction.
While the free-energy profile provides valuable insights into the reaction energetics, it does not provide direct information about the changes in the electron structure along the reaction coordinate.

\begin{figure}[t]
    \centering
    \includegraphics[width=0.9\linewidth]{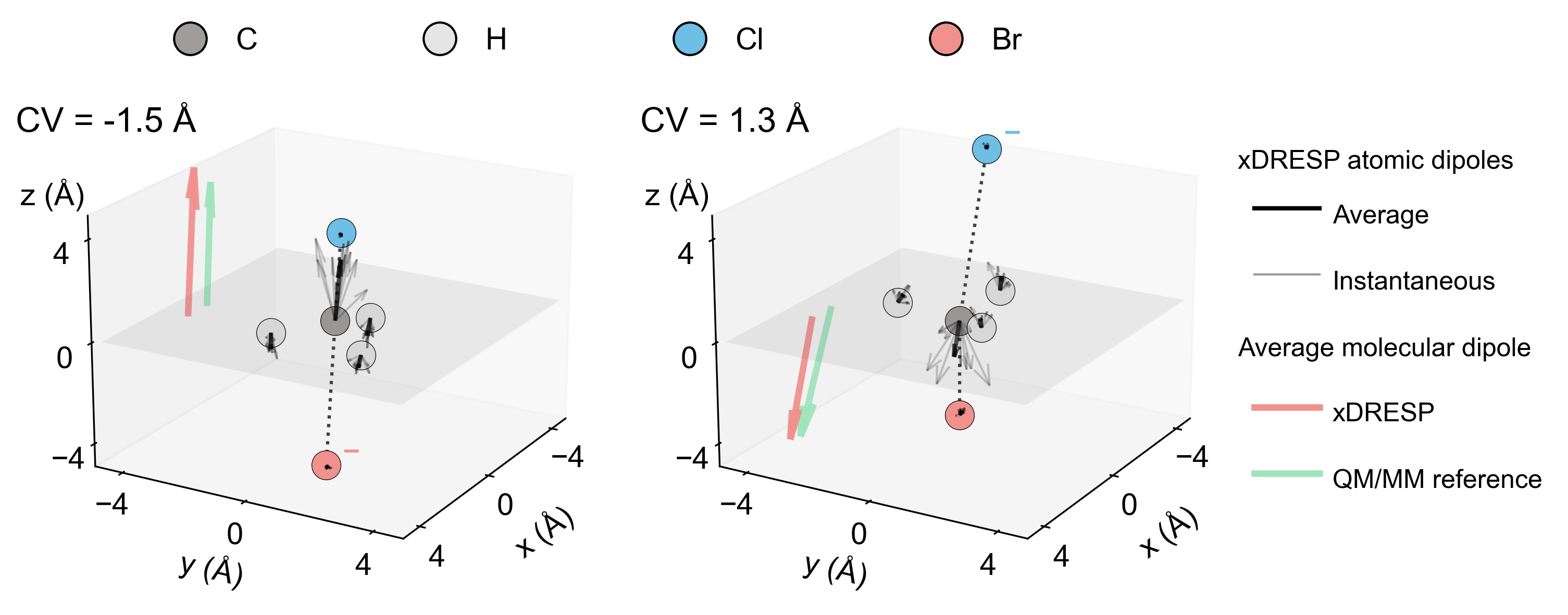}
    \caption{Three-dimensional representation of the xDRESP atomic dipole moments for the two minima in the free energy profile.
    For each TI window, the average coordinates for the atoms are represented, and the average xDRESP atomic dipoles are indicated for each atom with a black arrow. 
    In addition, instantaneous xDRESP dipoles from ten configurations during the MD are represented by gray arrows. 
    The average overall dipole moment of the QM subsystem is also reported, with a green arrow for the QM/MM reference and in red as calculated from the xDRESP atomic dipoles.
    }
    \label{fig:sn2_3d}
\end{figure}

To this end, we monitored the xDRESP charges and dipoles along the CV (Fig.~\ref{fig:sn2_results}).
Unsurprisingly, the most prominent changes in atomic charges occur for the \ce{Br} and \ce{Cl} atoms.
\ce{Br} starts as a negatively charged anion and, upon forming \ce{BrCH3}, its charge gets closer to zero, whereas \ce{Cl} undergoes the opposite process as it becomes the leaving anion.
The total xDRESP charge remains unchanged during the reaction, with its value about \num{-1.0}\,$e$.
These observations remain unchanged when the xDRESP dipoles are not fitted, i.e., only fitting xDRESP charges (Fig.~\ref{fig:SI_sn2_dresp}).
During the reaction, the presence of the negative anion on the opposite site of the \ce{CH3} group induces a flip in the overall molecular dipole of the QM region.
This is reflected in the xDRESP dipoles, and in particular for the \ce{C} and \ce{H} atoms whose atomic dipoles change during the reaction even though they remain nearly neutral in terms of atomic charges.
This behavior is more clearly visualized in three dimensions, as shown in Fig.~\ref{fig:sn2_3d}, where we report the average xDRESP dipoles for the two TI windows close to the free-energy minima. 

Finally, we evaluated the accuracy of the xDRESP fit in reproducing  $V^{\mathrm{QM \to MM}}$ as well as the overall molecular multipole moments (Figs.~\ref{fig:SI_sn2_potential} and~\ref{fig:SI_sn2_multipoles}).
As in the Phe--Br case, including atomic dipoles in the fit improves the accuracy in the description of the ESP. 
Although minor discrepancies remain, in this case the molecular dipoles and quadrupoles are also well captured, suggesting that additional restraints on higher-order atomic multipoles could improve the fit even further.

\section{\label{sec:conclusions}Conclusions and Outlook}
In this article, we introduced xDRESP, a scheme for fitting atom-centered electric multipole moments dynamically during QM/MM MD simulations.
In this approach, we extended the original D-RESP method by \citeauthor{laio2002dresp}\cite{laio2002dresp} beyond fitting the atomic point charges, to fit atom-centered multipoles of higher order. 
This approach is implemented in the MiMiC framework \cite{olsen2019mimic, antalik2024mimic}, which makes it automatically available for any combination of the interfaced QM and MM external programs. 

After validating the newly implemented scheme on a simple solute--solvent system, we proceeded to showcase three different aspects of the xDRESP approach.
We first focused on the fitting of atomic point charges for different systems, comparing the accuracy of the on-the-fly xDRESP model with fixed point-charge sets in commonly used FFs for biomolecular simulations.
For these systems, we observe a similar overall accuracy of QM/MM-generated xDRESP charges and classical FF, especially in reproducing the potential, $V^{\mathrm{QM \to MM}}$.
Even though the fixed point-charge FFs used for protein simulations performed well, the xDRESP charges and the fixed point-charge model based on averaged xDRESP charges outperform standard FFs for the description of the QM electrostatics in the case of nucleic acids, and especially generalized FFs for more chemically diverse ligands.
Afterwards, we focused on a halogenated system, for which we evaluated the effect of including higher-order atom-centered multipoles for an accurate description of the quantum ESP.
We observed that the inclusion of atomic dipoles greatly improves the accuracy of the model in terms of reproduction of the potential, albeit with some higher deviations in the molecular moments. 
This is mitigated by the inclusion of additional constraints to the overall charge and dipole moment, but the results suggest that the fitting procedure could still benefit from restraints applied to the higher-order atom-centered multipoles, similarly to the ones used in the fitting of atomic point charges.
Finally, we demonstrated the usefulness of the xDRESP approach as an on-the-fly analysis tool to monitor rearrangements of the charge density distribution by tracking the changes in the atom-center multipoles during an S$_\mathrm{N}$2 reaction.
It is interesting to note that the xDRESP analysis, including atomic dipoles, reveals differences between the Ph--Br and S$_\mathrm{N}$2 systems.
In the case of Ph--Br, the xDRESP point charges present large fluctuations that are somewhat reduced when xDRESP dipoles are included. 
However, these dipoles still show large fluctuations, especially for the aromatic carbon atoms. 
In contrast, for the S$_\mathrm{N}$2 system, the charge distribution evolves significantly along the reaction coordinate: the negatively charged halide ions modulate the dipoles of the central carbon atom and, to a lesser extent, those of the hydrogen atoms. 
This reflects induced polarization effects along the reaction rather than static anisotropy.

The results presented in this paper showcase possible applications of the xDRESP approach and serve as a stepping stone for future developments, e.g., in the context of force-matching approaches \cite{Maurer2007_FM, Doemer2014_FM, vona2025force_FM} to parametrize fixed point charge FFs, with possible extensions to polarizable FFs with atom-centered induced dipoles.
The inclusion of permanent atom-centered higher-order multipoles could also find application in the parameterization of more complex FFs, such as multipolar FFs \cite{jakobsen2015multipolar, Ponder2010AMOEBA}.
For this purpose, the approach might benefit from additional restraints on higher-order atom-centered multipoles, which would regularize the problem in a similar fashion as done for the derivation of atomic point charges.

Another possibility is the use of xDRESP charges and higher-order multipoles to define the interaction potentials between the QM and MM subsystems with an explicit dependence on the xDRESP multipoles.
This enables the implementation of an xDRESP-derived electrostatic embedding scheme, in which the MM atoms interact with the xDRESP multipoles instead of the QM density.
A similar three-layer electrostatic coupling scheme, though restricted to D-RESP point charges only, was earlier introduced by \citeauthor{laio2004variational} \cite{laio2004variational}: the MM atoms in the inner layer interact directly with the QM charge density, the ones in the middle layer with the D-RESP point charges, and the outer layer with the multipole expansion of the QM charge density, thus leading to a drastic reduction of the overall computational cost with only a minimal impact on the accuracy. 

\appendix
\section{\label{sec:eq_derivation}xDRESP least-squares problem}
In the xDRESP approach introduced in this work, the total loss-function for the least-squares problem (combining Eqs.~\ref{eq:leastsquare} and \ref{eq:leastsquare_constraints}) reads:
\begin{align}    
    L &= \sum_{p=1}^{N^\mathrm{SR}}{\left(V_p^{\mathrm{xDRESP}} - V_p^{\mathrm{QM \to MM}} \right)}^2 + w_R \sum_{i=1}^{N^\mathrm{QM}}{\left( M_i^{\mathrm{[0]}} - q_i^{\mathrm{ref}}\right)}^2 
    \nonumber\\
    &+\lambda_Q\left(\sum_{i=1}^{N^\mathrm{QM}} M_i^{\mathrm{[0]}} - Q_{\mathrm{tot}}^{\mathrm{QM}}\right) + \sum_{\xi\in\left\{x,y,z\right\}}  \sum_{\left|\gamma\right|=1} \lambda_{D, \xi} \;\gamma_\xi \left[\sum_{i=1}^{N^{\mathrm{QM}}}\left( M_i^{\left[0\right]}R_{i, \xi} + M_{i}^{\left[\gamma\right]}\right) - D^{\mathrm{QM}}_{\mathrm{tot, \xi}}\right],
\end{align}
and needs to be minimized with respect to the atom-centered multipole moments and the Lagrange multipliers, i.e., 
\begin{equation}
    \begin{cases}
    \frac{\partial L}{\partial M_j^{\left[\beta\right]}} &= 0 \quad\forall j=1,\dots,N^{\mathrm{QM}},\,\, \left|\beta\right|=0,\dots,\Lambda\\
    \frac{\partial L}{\lambda_\sigma^{\left[\beta\right]}} &=0 \quad \mathrm{for }\,\,\sigma\in \{Q, D_x, D_y, D_z\}
    \end{cases}
\end{equation}
\noindent
The minimization with respect to the multipole moments leads to
\begin{align}
    0 &= \sum_{p=1}^{N^\mathrm{SR}}{\left(V_p^{\mathrm{xDRESP}} - V_p^{\mathrm{QM \to MM}} \right)} \frac{\partial V_p^{\mathrm{xDRESP}}}{\partial M_j^{\left[\beta\right]}} 
    +  w_R \sum_{i=1}^{N^\mathrm{QM}}{\left( M_i^{\mathrm{[0]}} - q_i^{\mathrm{ref}}\right)} \frac{\partial M_i^{\mathrm{[0]}}}{\partial M_j^{\left[\beta\right]}} \nonumber\\
    &\quad+\frac{1}{2} \lambda_Q \sum_{i=1}^{N^\mathrm{QM}}     \frac{\partial M_i^{\mathrm{[0]}}}{\partial M_j^{\left[\beta\right]}} 
    + \frac{1}{2} \sum_{\xi\in\left\{x,y,z\right\}}  \sum_{\left|\gamma\right|=1} \lambda_{D, \xi} \;\gamma_\xi \sum_{i=1}^{N^{\mathrm{QM}}}\left( \frac{\partial M_i^{\left[0\right]}}{\partial M_j^{\left[\beta\right]}} R_{i, \xi} 
    + \frac{\partial M_{i}^{\left[\gamma\right]}}{\partial M_j^{\left[\beta\right]}}\right)\nonumber\\
    &= \sum_{p=1}^{N^\mathrm{SR}}{\left(V_p^{\mathrm{xDRESP}} - V_p^{\mathrm{QM \to MM}} \right)} \frac{\left(-1\right)^{\left|\beta\right|}}{\beta!} T^{\left[\beta\right]}{\left(\mathbf{R}_p, \mathbf{R}_j\right)} 
    +  w_R \delta_{\beta,0}\delta_{i,j}\left( M_j^{\mathrm{[0]}} - q_j^{\mathrm{ref}}\right)  \nonumber\\
    &\quad+\frac{1}{2} \lambda_Q \delta_{\beta,0}\delta_{i,j}
    + \frac{1}{2} \sum_{\xi\in\left\{x,y,z\right\}}  \sum_{\left|\gamma\right|=1} \lambda_{D, \xi} \;\gamma_\xi \left( \delta_{\beta, 0} \delta_{i, j} R_{i, \xi} + \delta_{\beta,\gamma}\delta_{i,j}\right),
\end{align}
and using the definition of $V_p^{\mathrm{xDRESP}}$ from Eq.~\ref{eq:dresp_pot}, it is possible to rearrange the equation as
\begin{align}\label{eq:minimization_multipoles}
    &\sum_{i=1}^{N^{\mathrm{QM}}} \sum_{\left|\alpha\right|=0}^{\Lambda}\left[ \sum_{p=1}^{N^\mathrm{SR}} \frac{\left(-1\right)^{\left|\alpha\right| + \left|\beta\right|}}{\alpha!\;\beta!} 
T^{\left[\alpha\right]}{\left(\mathbf{R}_p, \mathbf{R}_i\right)}
T^{\left[\beta\right]}{\left(\mathbf{R}_p, \mathbf{R}_j\right)} +  w_R \delta_{\beta, 0} \delta_{i, j}
    \right]
    M_i^{\left[\alpha\right]} \nonumber\\
    &\quad+ \frac{1}{2}\lambda_Q \delta_{\beta,0}\delta_{i,j} 
    + \frac{1}{2} \sum_{\xi\in\left\{x,y,z\right\}}  \sum_{\left|\gamma\right|=1} \lambda_{D, \xi} \;\gamma_\xi   \left(\delta_{\beta,0}\delta_{i,j} R_{j, \xi} +\delta_{\beta,\gamma}\delta_{i,j}\right)  \nonumber\\
    &= \sum_{p=1}^{N^\mathrm{SR}} \frac{\left(-1\right)^{\left|\beta\right|}}{\beta!}  V_p^{\mathrm{QM \to MM}} T^{\left[\beta\right]}{\left(\mathbf{R}_p, \mathbf{R}_j\right)}  +  w_R \delta_{\beta, 0} \delta_{i,j}q_j^{\mathrm{ref}},
\end{align}
which can be reduced to 
\begin{equation}
\sum_{i=1}^{N^{\mathrm{QM}}}\left[ \sum_{\left|\alpha\right|=0}^{\Lambda} A_{i,j}^{\left[\alpha\right]\left[\beta\right]}  M_i^{\left[\alpha\right]} + \frac{1}{2} \lambda_Q \delta_{\beta,0}\delta_{i,j} + \frac{1}{2} \sum_{\xi\in\left\{x,y,z\right\}}  \sum_{\left|\gamma\right|=1} \lambda_{D, \xi} \;\gamma_\xi \left( \delta_{\beta, 0} \delta_{i,j} R_{j, \xi} + \delta_{\beta,\gamma}\delta_{i,j}\right) \right] = b_j^{\left[\beta\right]}.
\end{equation}
\noindent
The minimization with respect to the Lagrange multipliers simply recovers the constraints
\begin{equation}\label{eq:minimization_charge}
\sum_{i=1}^{N^\mathrm{QM}} M_i^{\mathrm{[0]}} = Q_{\mathrm{tot}}^{\mathrm{QM}},
\end{equation}
\begin{equation}\label{eq:minimization_dipole}
\sum_{i=1}^{N^{\mathrm{QM}}}\left( M_i^{\left[0\right]}R_{i, \xi} + M_{i}^{\left[\gamma\right]}\right) = D^{\mathrm{QM}}_{\mathrm{tot, \xi}}\quad\mathrm{for}\quad\xi\in\left\{x,y,z \right\}, \;\mathrm{with}\left\{\begin{smallmatrix}
    \gamma_\xi&=1\\
    \left|\gamma\right|&=1
\end{smallmatrix}\right..    
\end{equation}
\noindent
Combining Eqs.~\ref{eq:minimization_multipoles}, \ref{eq:minimization_charge}, and \ref{eq:minimization_dipole} leads to the system of equations reported in the main text (Eq.~\ref{eq:matrix_problem}).

\section*{Author Information}
\subsection*{Author Contributions}
\textbf{AL}: conceptualization (equal), data curation (lead), investigation (lead), software (lead), visualization (lead), writing — original draft (lead), writing — review \& editing (equal).
\textbf{AA}: conceptualization (equal), software (supporting), investigation (supporting), visualization (supporting), writing — review \& editing (equal).
\textbf{JMHO}: conceptualization (equal), funding acquisition (supporting), supervision (equal), writing — review \& editing (equal).
\textbf{UR}: conceptualization (equal), funding acquisition (lead), supervision (equal), writing — review \& editing (equal).
\subsection*{Notes}
The authors declare no competing financial interest.

\section*{Acknowledgments}
We gratefully acknowledge Thibault Kl\"ay for his support in the QM/MM equilibration of the S$_\mathrm{N}$2 system.
This work has been supported by the Swiss National Science Foundation (grant 200020-185092 and 200020-219440), and used computing time from the Swiss National Computing Centre CSCS.
JMHO gratefully acknowledges financial support from VILLUM FONDEN (Grant No.~VIL29478).

\section*{Data and Software Availability}
The source code of the MiMiC framework is free and open-source and is hosted on GitLab \cite{mimic-projects}. 
The xDRESP implementation will be part of the upcoming release of MiMiC. 
The version of MiMiC used within this study, together with all the data generated, is available on Zenodo under \url{https://zenodo.org/records/17521583}, including a Jupyter Notebook to reproduce the analysis and generate the plots and tables presented.

\begin{suppinfo}
Supporting Information contains details on all the systems and parameters, as well as additional figures and tables not included in the main text.
\end{suppinfo}

{\footnotesize
\bibliography{bib}
}

\newpage

\section*{Supporting Information}
\setcounter{figure}{0} 
\setcounter{table}{0} 
\renewcommand{\thetable}{S\arabic{table}}
\renewcommand{\thefigure}{S\arabic{figure}}

\subsection*{Detailed list of computational parameters}

\begin{table}[h!]
\centering
\caption{MM subsystem parameters: total number of atoms and sizes of the simulation boxes used. 
For all the systems, cubic boxes have been used.}
\label{tab:mm_params}
\begin{tabular}{l|c|c}
System    & Atoms Number   & Box Size (nm)           \\ \hline
Ace   & \num{2934}   & \num{3.11}  \\
AlaGly   & \num{10653}   & \num{4.76}  \\
Gua   & \num{23083}   & \num{6.13}  \\
CREB--APAP   & \num{105294}   & \num{10.21}  \\
Ph--Br  & \num{16278}   & \num{5.49}  \\
S$_\mathrm{N}$2  & \num{81140}   & \num{9.94}  \\
\end{tabular}
\end{table}

\begin{table}[h!]
\centering
\caption{QM subsystem parameters: total number of atoms and lists of QM parameters used.
For QM subsystems treated with CPMD, plane-wave (PW) cutoffs are reported, and for CP2K, density cutoffs and relative cutoffs. 
For all the systems, cubic boxes have been used, and the box sizes are also reported.
}
\label{tab:qm_params}
\begin{tabular}{l|c|ccccc}
\makecell{System} & \makecell{Atoms\\Number} 
    &  \makecell{PW Cutoff\\(Ry)} 
   & \makecell{Density Cutoff\\(Ry)} 
   & \makecell{Rel. Cutoff\\(Ry)} 
   & \makecell{Box Size\\(\si{\angstrom})} \\ 
\hline
Ace     & \num{10}  & 80   & 450  & 50  &  10.0 \\
AlaGly  & \num{20}  & --   & 200 & 90  &  20.0 \\
Gua  & \num{37}  & --   & 320 & 110  &  30.0 \\
CREB--APAP  & \num{20}  & --   & 300 & 70  &  20.0 \\
Ph--Br  & \num{12}  & --   & 260 & 80 & 20.0 \\
S$_\mathrm{N}$2  & \num{6}  &  \num{130} & -- & -- & 25.0 \\
\end{tabular}%
\end{table}
\null

\begin{table}[h!]
\centering
\caption{QM/MM parameters for MiMiC electrostatic coupling: short-range cutoff, and multipole expansion order of the QM subsystem charge density.
The sorting between short and long ranges for the MM atoms has been performed every 50 MD steps.}
\label{tab:mimic_params}
\begin{tabular}{l|c|c}
System    & SR cutoff ($a_0$)   & Multipole order            \\ \hline
Ace     & 30    & 5 \\
AlaGly   & 30   & 7  \\
Gua   & 45   & 7  \\
CREB--APAP   & 30   & 5  \\
Ph--Br   & 25   & 5  \\
S$_\mathrm{N}$2 & 30   & 9  \\
\end{tabular}
\end{table}
\clearpage

\subsection*{Validation of the xDRESP approach}
\subsubsection*{Ace system}

\begin{table}[ht]
\centering
\begin{tabular}{c | c c c c c}
$\mathbf{w}_\mathrm{R}$ & O & \ce{C=O} & \ce{C-H3} & \multicolumn{2}{c}{H} \\
\hline
$\mathbf{w}_\mathrm{R}\,\,0.0$ & $-0.68 \pm 0.04$ & $0.81 \pm 0.15$ & $-0.57 \pm 0.36$ & $0.17 \pm 0.09$ & $0.15 \pm 0.09$ \\
 &  &  & $-0.49 \pm 0.34$ & $0.15 \pm 0.09$ & $0.15 \pm 0.10$ \\
 &  &  &  & $0.18 \pm 0.09$ & $0.14 \pm 0.10$ \\
\hline
$\mathbf{w}_\mathrm{R}\,\,10^{-5}$ & $-0.67 \pm 0.04$ & $0.72 \pm 0.11$ & $-0.43 \pm 0.29$ & $0.13 \pm 0.07$ & $0.11 \pm 0.08$ \\
 &  &  & $-0.35 \pm 0.28$ & $0.12 \pm 0.07$ & $0.11 \pm 0.08$ \\
 &  &  &  & $0.14 \pm 0.07$ & $0.11 \pm 0.08$ \\
\hline
$\mathbf{w}_\mathrm{R}\,\,10^{-4}$ & $-0.64 \pm 0.04$ & $0.52 \pm 0.07$ & $-0.13 \pm 0.12$ & $0.06 \pm 0.04$ & $0.06 \pm 0.05$ \\
 &  &  & $-0.09 \pm 0.13$ & $0.05 \pm 0.04$ & $0.06 \pm 0.04$ \\
 &  &  &  & $0.08 \pm 0.04$ & $0.05 \pm 0.05$ \\
\hline
$\mathbf{w}_\mathrm{R}\,\,10^{-3}$ & $-0.55 \pm 0.04$ & $0.30 \pm 0.05$ & $-0.02 \pm 0.02$ & $0.05 \pm 0.02$ & $0.05 \pm 0.03$ \\
 &  &  & $-0.02 \pm 0.03$ & $0.03 \pm 0.03$ & $0.05 \pm 0.03$ \\
 &  &  &  & $0.07 \pm 0.03$ & $0.04 \pm 0.03$ \\
\hline
$\mathbf{w}_\mathrm{R}\,\,10^{-2}$ & $-0.44 \pm 0.04$ & $0.13 \pm 0.02$ & $-0.04 \pm 0.01$ & $0.08 \pm 0.02$ & $0.06 \pm 0.03$ \\
 &  &  & $-0.04 \pm 0.01$ & $0.04 \pm 0.03$ & $0.06 \pm 0.03$ \\
 &  &  &  & $0.08 \pm 0.03$ & $0.08 \pm 0.02$ \\
\hline
$\mathbf{w}_\mathrm{R}\,\,10^{-1}$ & $-0.35 \pm 0.04$ & $0.11 \pm 0.01$ & $-0.05 \pm 0.01$ & $0.08 \pm 0.02$ & $0.04 \pm 0.03$ \\
 &  &  & $-0.05 \pm 0.01$ & $0.03 \pm 0.04$ & $0.04 \pm 0.03$ \\
 &  &  &  & $0.06 \pm 0.03$ & $0.09 \pm 0.01$ \\
\hline
$\mathbf{w}_\mathrm{R}\,\,1.0$ & $-0.26 \pm 0.03$ & $0.13 \pm 0.01$ & $-0.06 \pm 0.01$ & $0.05 \pm 0.01$ & $0.03 \pm 0.02$ \\
 &  &  & $-0.06 \pm 0.00$ & $0.03 \pm 0.03$ & $0.04 \pm 0.02$ \\
 &  &  &  & $0.04 \pm 0.02$ & $0.06 \pm 0.01$ \\
\hline
$\mathbf{Hirshfeld}$ & $-0.23 \pm 0.02$ & $0.14 \pm 0.01$ & $-0.06 \pm 0.01$ & $0.04 \pm 0.01$ & $0.03 \pm 0.01$ \\
 &  &  & $-0.06 \pm 0.00$ & $0.03 \pm 0.02$ & $0.03 \pm 0.01$ \\
 &  &  &  & $0.04 \pm 0.01$ & $0.04 \pm 0.01$ \\
\hline
\end{tabular}
\caption{Average xDRESP point charges during a \SI{1}{\pico\second} QM/MM MD simulation of acetone in water obtained with different $w_\mathrm{R}$ restraint values.
The error reported corresponds to the standard deviation.}
\label{tab:act_dresp}
\end{table}

\begin{figure}[h]
    \centering
    \includegraphics[width=\linewidth]{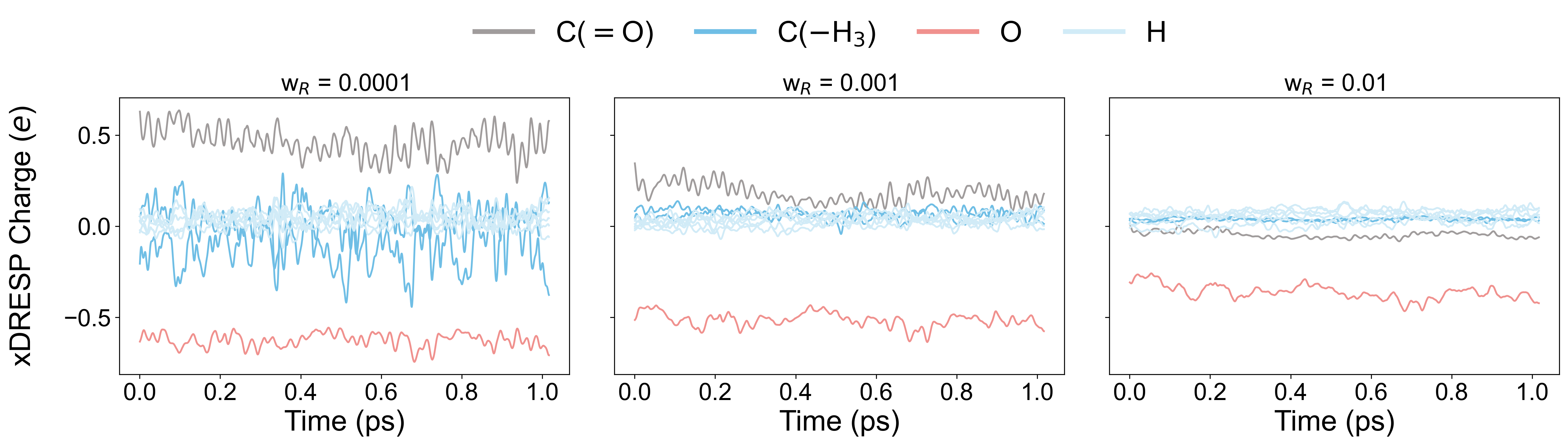}
    \caption{xDRESP charges during a \SI{1}{\pico\second} QM/MM MD simulation of acetone in water obtained with $w_\mathrm{R}$ = \num{e-4}--\num{e-2} and zeroes as reference charges.}
    \label{fig:SI_acetone_ref0}
\end{figure}

\begin{figure}[h]
    \centering
    \includegraphics[width=.6\linewidth]{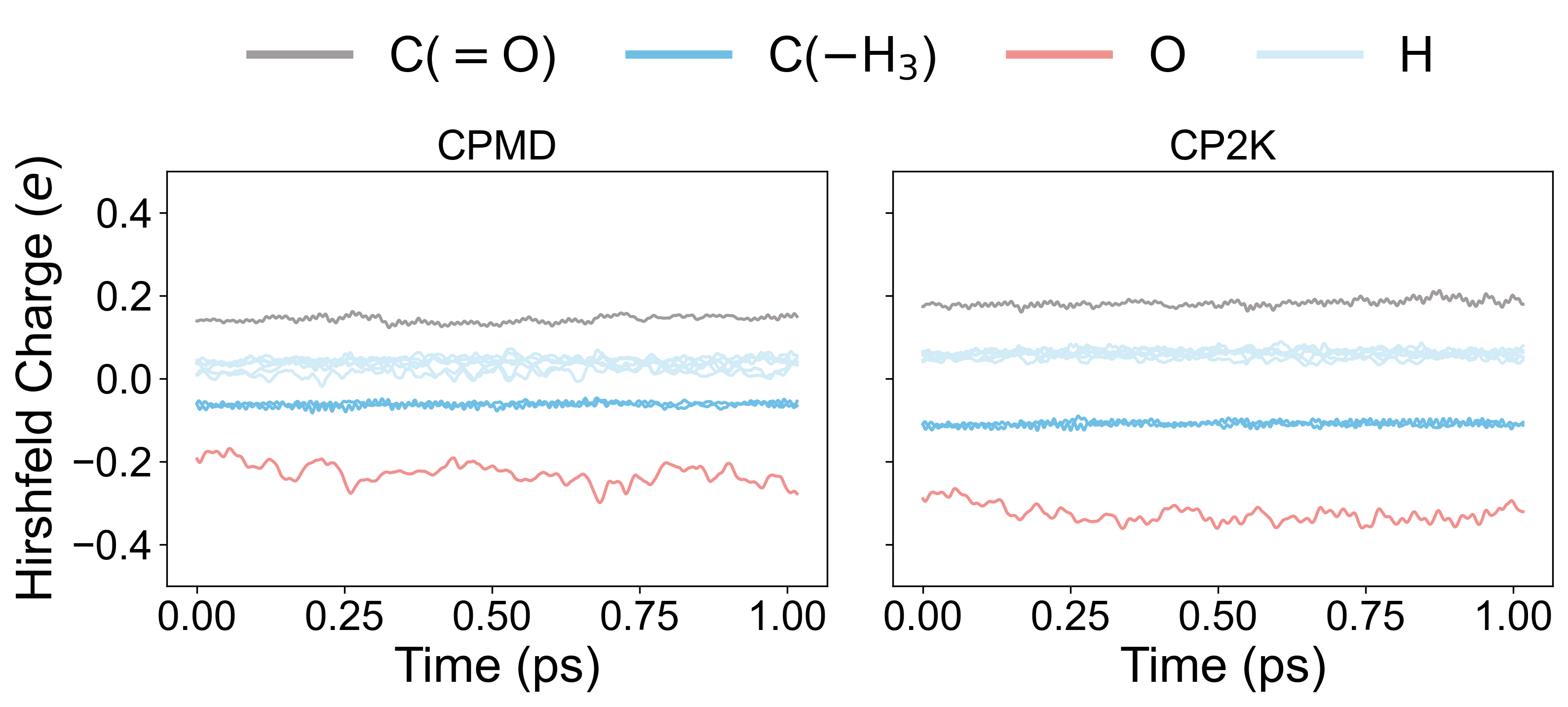}
    \caption{Comparison of Hirshfeld charges from CPMD and CP2k during a \SI{1}{\pico\second} QM/MM MD simulation of acetone in water.}
    \label{fig:SI_hirshfeld}
\end{figure}

\begin{figure}
    \centering
    \includegraphics[width=.6\linewidth]{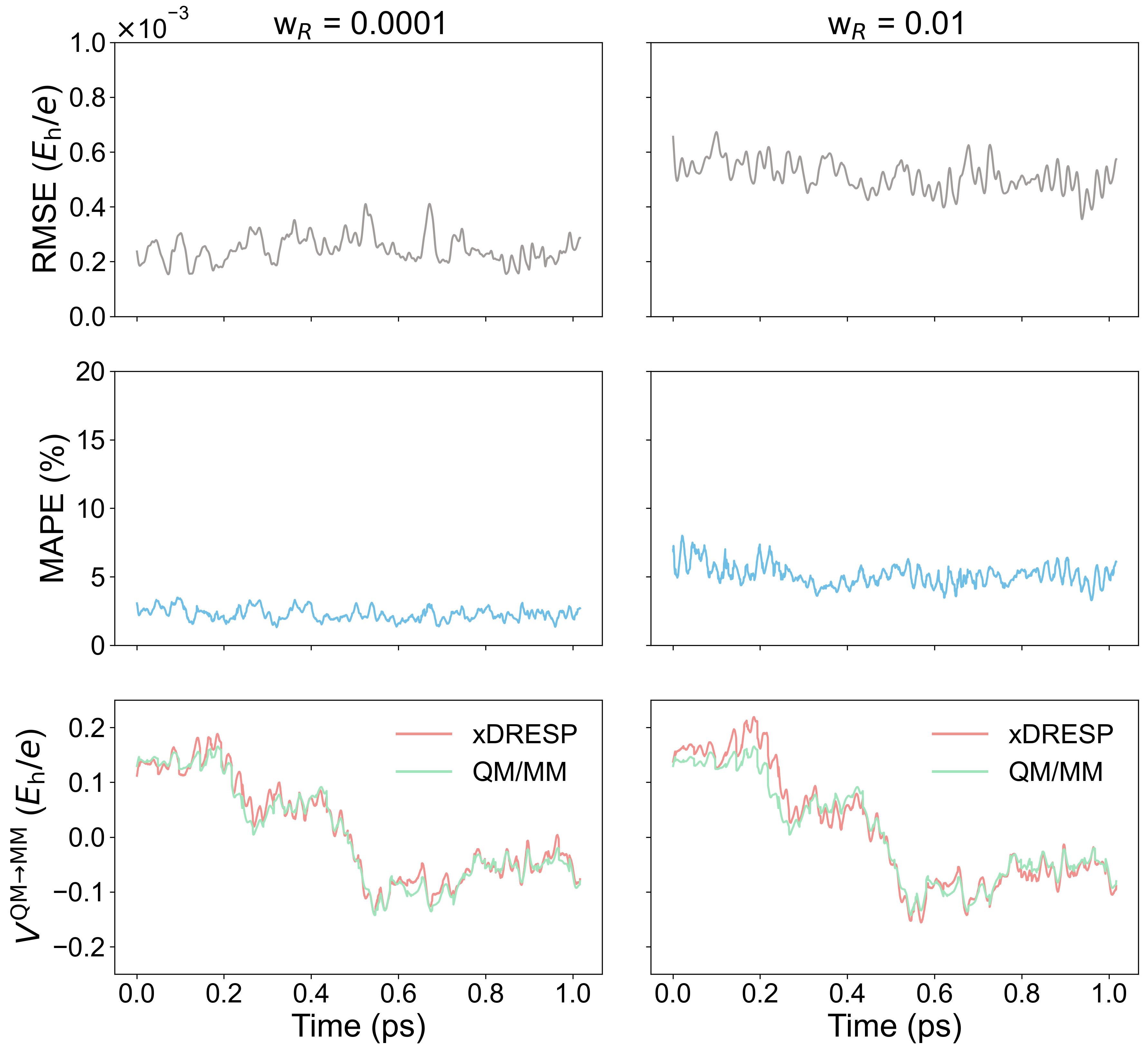}
    \caption{Different metrics used to assess the accuracy of the xDRESP point charge set obtained with $w_\mathrm{R}$ = \num{e-4} and \num{e-2} in reproducing $V^{\mathrm{QM \to MM}}$.}
    \label{fig:SI_act_dresp_metrics_pot}
\end{figure}

\begin{figure}
    \centering
    \includegraphics[width=.6\linewidth]{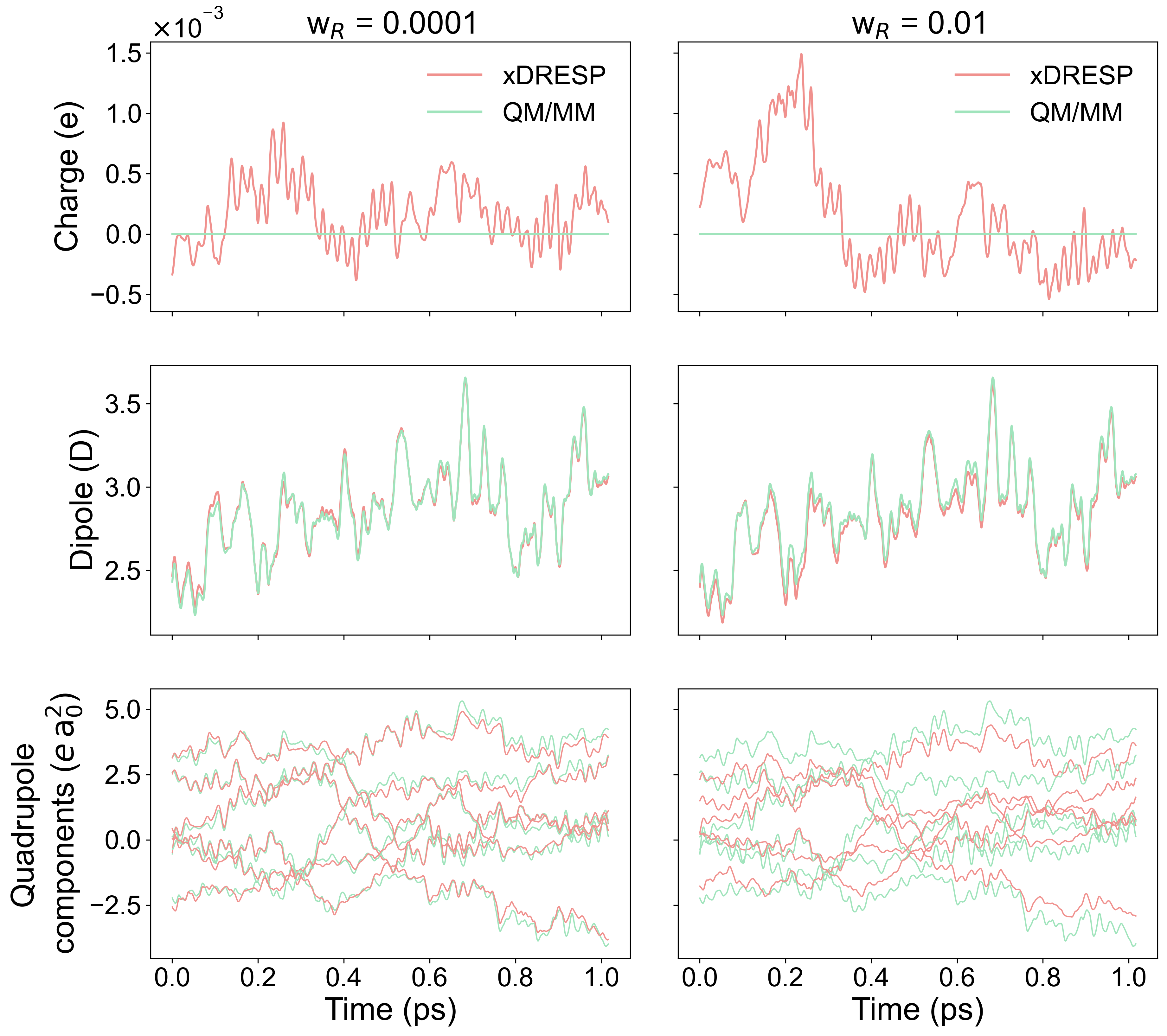}
    \caption{Different metrics used to assess the accuracy of the xDRESP point charge set obtained with $w_\mathrm{R}$ = \num{e-4} and \num{e-2} in reproducing the molecular multipoles.}
    \label{fig:SI_act_dresp_metrics_multipoles}
\end{figure}

\begin{figure}
    \centering
    \includegraphics[width=.6\linewidth]{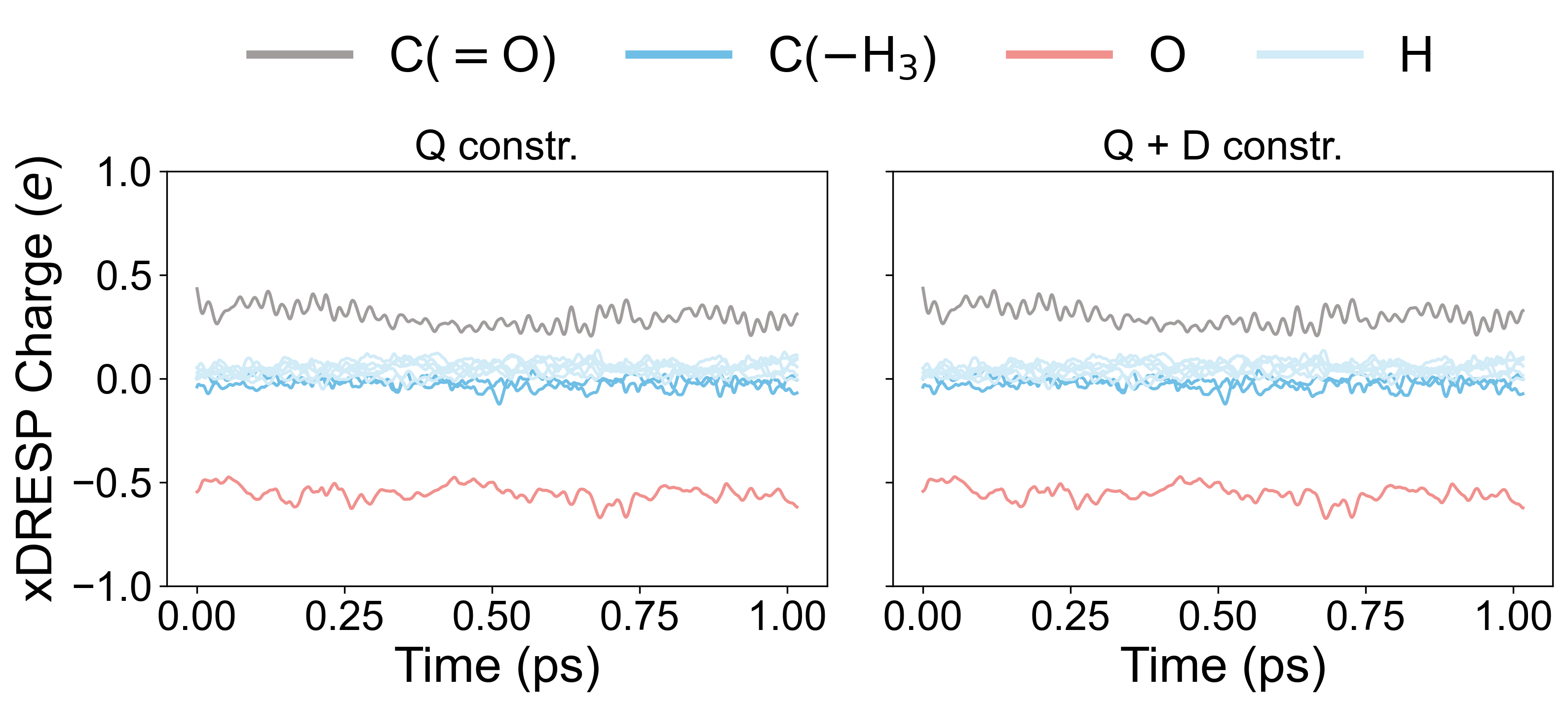}
    \caption{Comparison of the xDRESP charges during a \SI{1}{\pico\second} QM/MM MD simulation of acetone in water with the additional constraints on the total charge (Q const.) and on the charge and the dipole components (Q+D constr.).}
    \label{fig:SI_constr_dresp}
\end{figure}

\begin{figure}
    \centering
    \includegraphics[width=.6\linewidth]{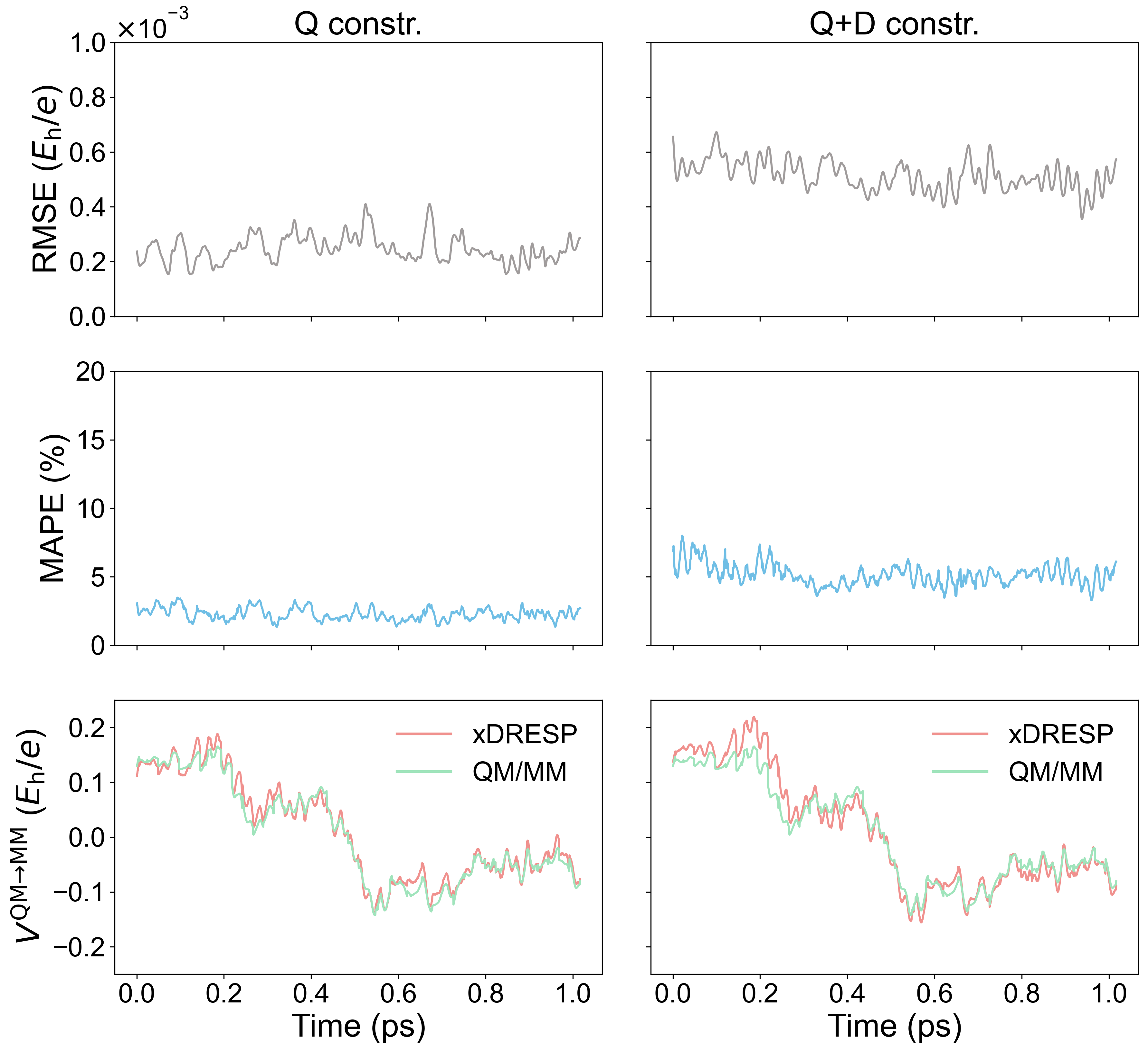}
    \caption{Different metrics used to assess the accuracy of the xDRESP point charge set obtained with constraints on the total charge (Q constr.) or on the charge and the dipole (Q+D constr.), in reproducing $V^{\mathrm{QM \to MM}}$.}
    \label{fig:SI_constr_dresp_metrics_potential}
\end{figure}

\begin{figure}
    \centering
    \includegraphics[width=.6\linewidth]{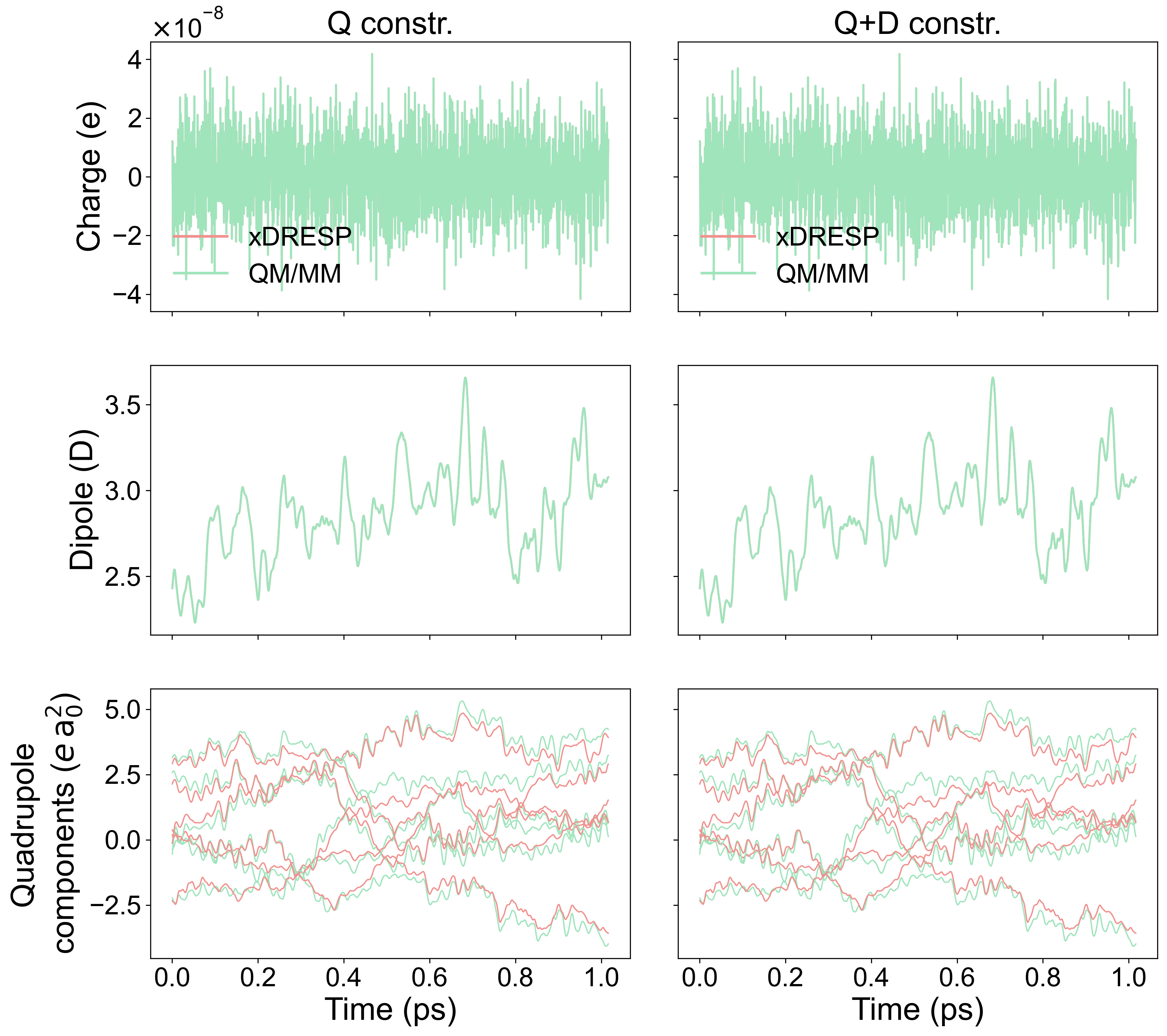}
    \caption{Different metrics used to assess the accuracy of the xDRESP point charge set obtained with constraints on the total charge (Q constr.) or on the charge and the dipole (Q+D constr.), in reproducing the molecular multipoles.
    When not visible, the xDRESP line is overlapped by the QM/MM reference.}
    \label{fig:SI_constr_dresp_metrics_multipoles}
\end{figure}

\null
\clearpage
\subsection*{Point charge models for classical force fields}

\subsubsection*{AlaGly system}

\begin{figure}[ht!]
    \centering
    \includegraphics[width=.7\linewidth]{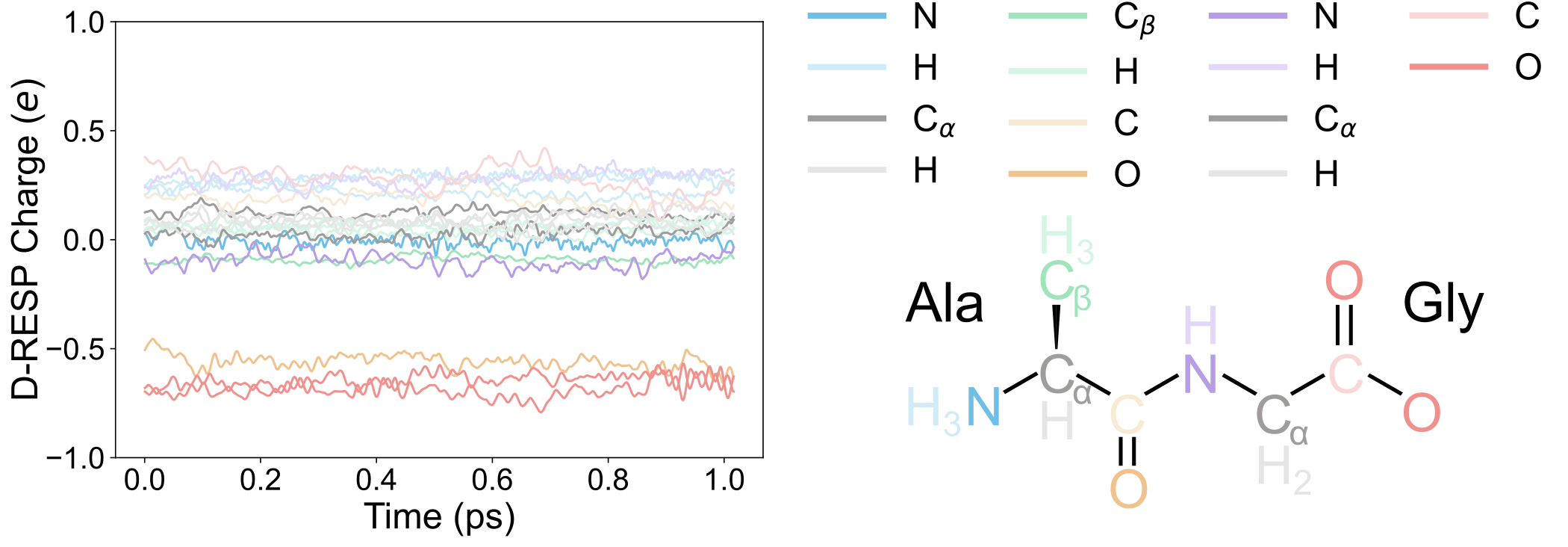}
    \caption{xDRESP charges for the AlaGly dipeptide during \SI{1}{\pico\second} QM/MM MD with a restraint to the Hirshfeld charges $w_\textrm{R}$=\num{e-3}.\vspace{-.5cm}}
    \label{fig:SI_alagly_dresp_charges}
\end{figure}

\begin{table}[hb!]
\centering
\resizebox{.85\textwidth}{!}{%
\begin{tabular}{c c| S S S S | c | c }
Atom & Residue & {AMBER} & {CHARMM}  & {OPLS-AA/L} & {GROMOS}  & {$\left<\right.$Hirshfeld$\left.\right>$ } & {$\left<\right.$xDRESP$\left.\right>$ }  \\
\hline
N  & Ala 
    & 0.1414 & -0.30 & -0.30 & 0.129 & 0.024 & -0.007\\
\hline
H & Ala 
    & 0.1997 & 0.33 & 0.33 & 0.248 & \makecell{0.215 \\ 0.216 \\ 0.221} & \makecell{0.220 \\ 0.267 \\ 0.284} \\
\hline    
C$_\alpha$ & Ala 
    & 0.0962 & 0.21 & 0.25 & 0.127 & 0.065 &  0.124 \\
\hline
H$_\alpha$ & Ala  
    & 0.0889 & 0.10 & 0.06 & // & 0.088 & 0.091 \\
\hline
C$_\beta$ & Ala 
    & -0.0597 & -0.27 & -0.18 & 0.000 & -0.089& -0.092 \\
\hline
H$_\beta$ & Ala 
    & 0.0300 & 0.09 & 0.06  & // & \makecell{0.063 \\ 0.069 \\ 0.064} & \makecell{0.066 \\ 0.044 \\ 0.055} \\
\hline
C & Ala 
    & 0.6163 & 0.51  & 0.50 & 0.450 & 0.170 & 0.188\\
\hline
O & Ala 
    & -0.5722 & -0.51  & -0.50  & -0.450 & -0.341 & -0.560\\
\hline
N & Gly 
    & -0.3821 &  -0.47 & -0.50  & -0.310 & -0.074 & -0.098 \\
\hline
H & Gly 
    & 0.2681 & 0.31 & 0.30   & 0.310 & 0.171 & 0.282 \\
\hline
C$_\alpha$ & Gly 
    & -0.2493 & -0.02 & -0.02 & 0.000 & -0.033 & 0.027  \\
\hline
H$_\alpha$ & Gly 
    & 0.1056 & 0.09 & 0.06 & // & \makecell{0.053 \\ 0.037 } & \makecell{0.073 \\ 0.089 }  \\
\hline
C & Gly 
    & 0.7231 & 0.34 & 0.70 & 0.270 & 0.097 & 0.283  \\
\hline
O & Gly 
    & -0.7855 & -0.67 & -0.80 & -0.635 &  \makecell{-0.518 \\ -0.499 } & \makecell{-0.683 \\ -0.651 }  \\
\end{tabular}
}
\caption{Point charges for AlaGly dipeptide from common biomolecular force fields and xDRESP fit.
Atom names refer to the scheme in Fig.~\ref{fig:SI_alagly_dresp_charges}.
For the AMBER force field, the charges present the same values in FF14SB, FF19SB, and FF99SB. 
For the CHARMM force field, the charges present the same values in CHARMM27 (CHARM22 with CMAP for proteins) and CHARMM36.
GROMOS refers to GROMOS 54A7 (united force field, for which the charges of atoms not present in the model are indicated by `//').
The values of  $\left<\right.$Hirshfeld$\left.\right>$  and $\left<\right.$xDRESP$\left.\right>$ correspond to the average during MD.}
\label{tab:SI_alagly_point_charges}
\end{table}

\begin{figure}
    \centering
    \includegraphics[width=.8\linewidth]{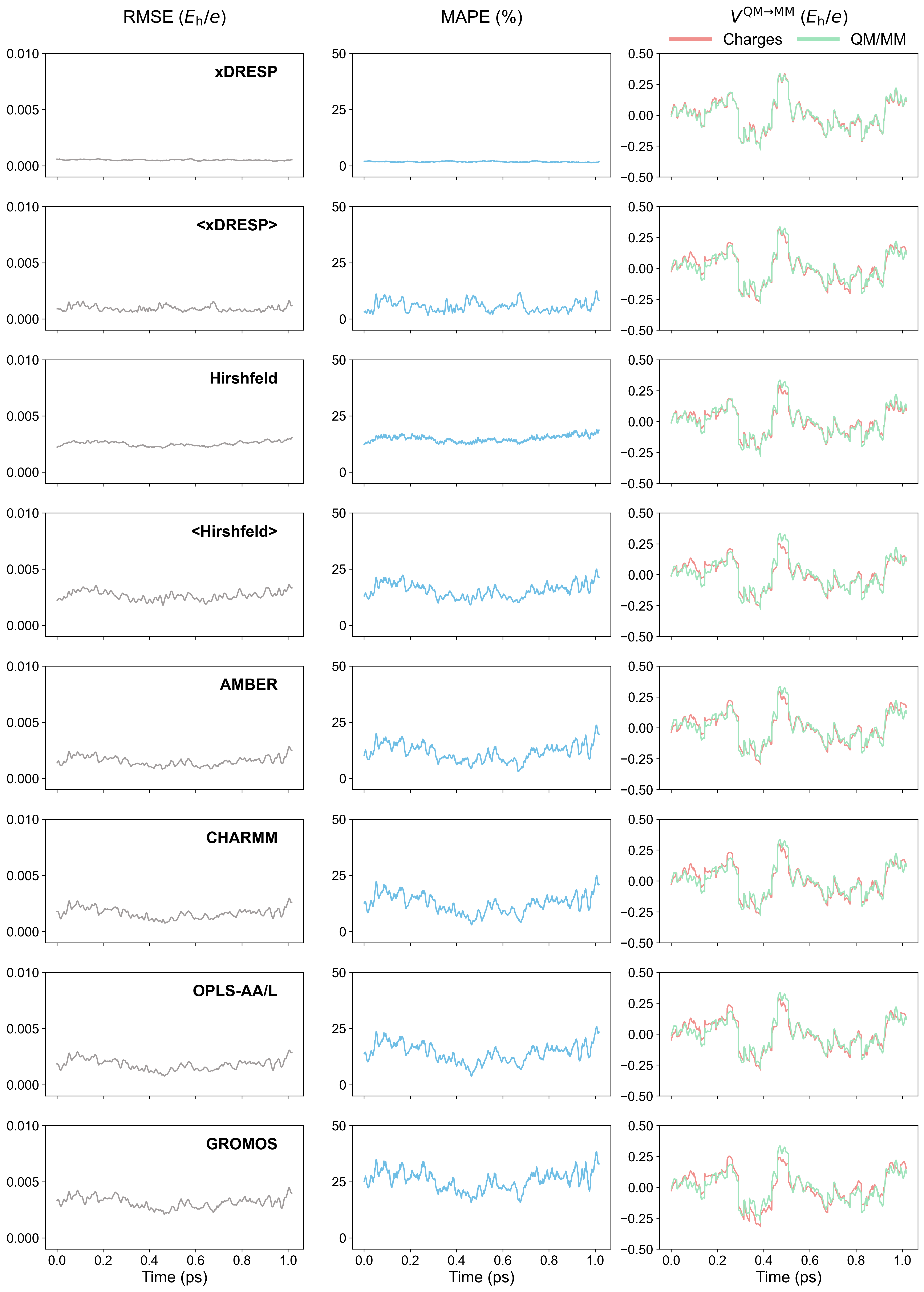}
    \caption{Comparison between xDRESP charges fitted at every MD step and different point charge models in reproducing $V^{\mathrm{QM \to MM}}$ for the AlaGly system. 
    The potential is given relative to the average potential ($\Delta V^{\mathrm{QM \to MM}}(t) = V^{\mathrm{QM \to MM}}(t) - \langle V^{\mathrm{QM \to MM}}\rangle$).}
    \label{fig:SI_alagy_metric_potential}
\end{figure}

\begin{figure}
    \centering
    \includegraphics[width=.8\linewidth]{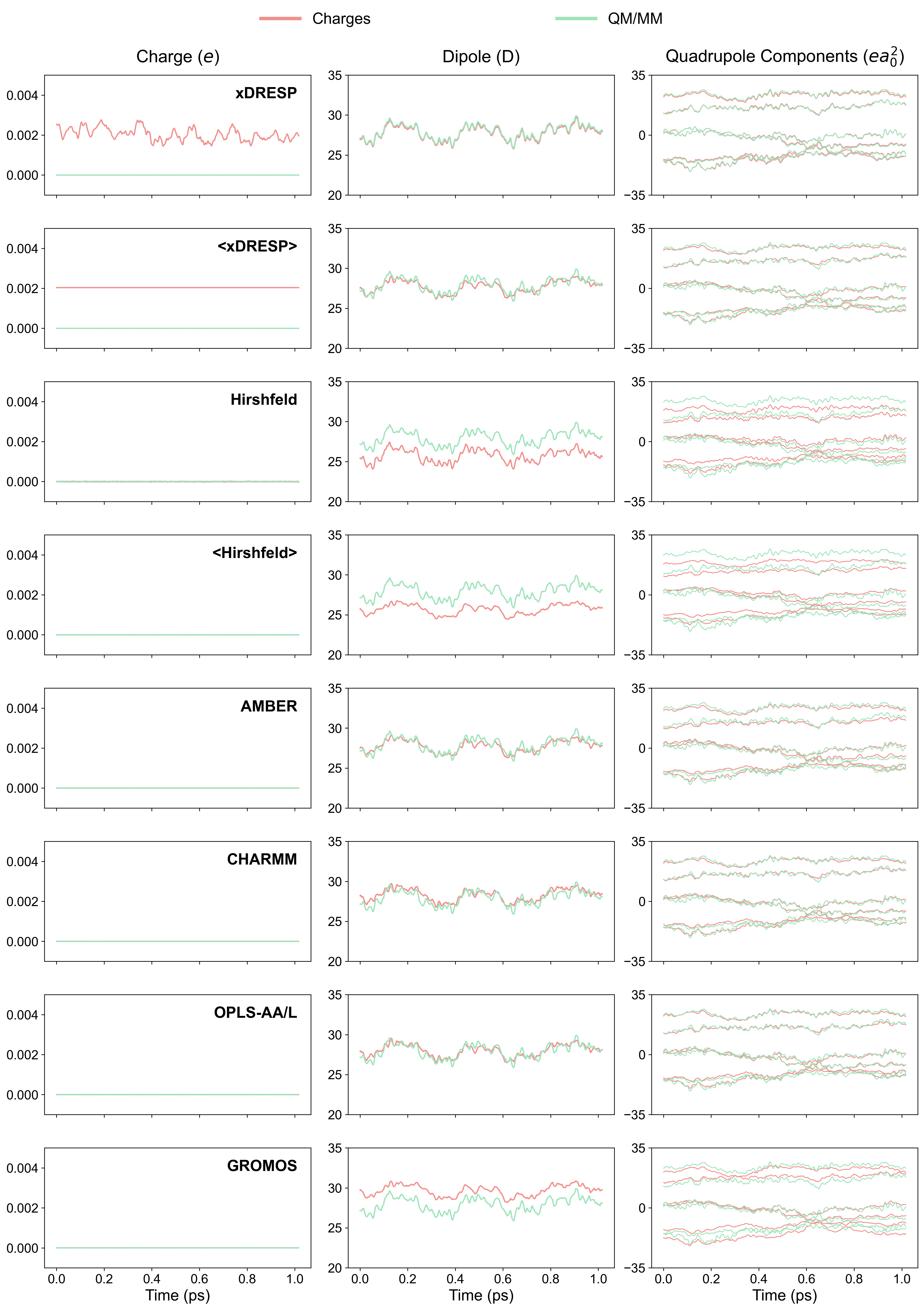}
    \caption{Comparison between xDRESP charges fitted at every MD step and different point charge models in reproducing the molecular multipoles from QM/MM for the AlaGly system. 
    When not visible, the line corresponding to the point charge set (red) is overlapped by the QM/MM reference (green).}
    \label{fig:SI_alagy_metric_multipoles}
\end{figure}

\newpage
\subsubsection*{Gua system}
\begin{figure}
    \centering
    \includegraphics[width=.9\linewidth]{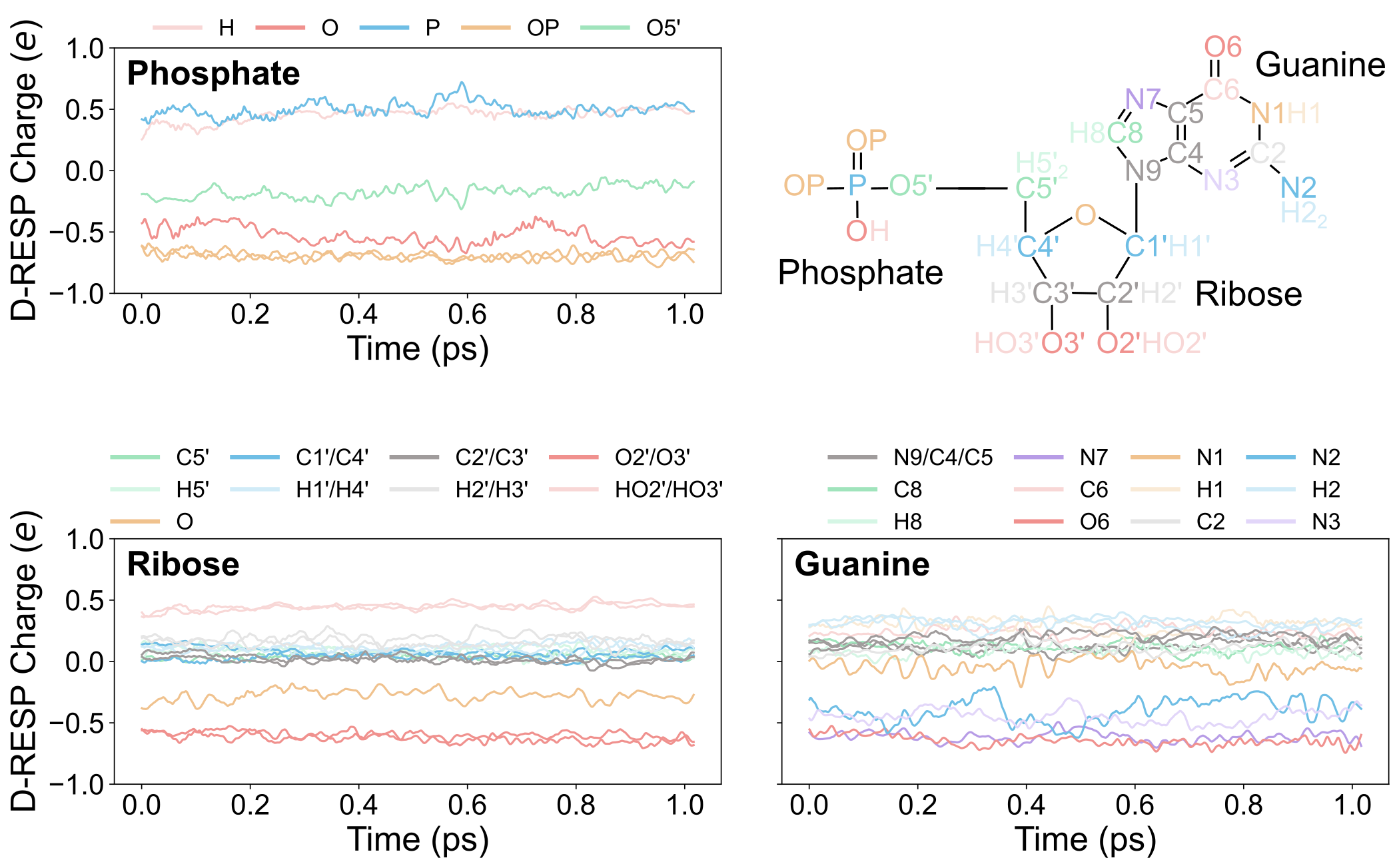}
    \caption{xDRESP charges for the Gua nucleotide during \SI{1}{\pico\second} QM/MM MD with a restraint to the Hirshfeld charges $w_\textrm{R}$=\num{e-3}.}
    \label{fig:SI_gua_dresp_charges}
\end{figure}

\begin{table}
\centering
\resizebox{.9\textwidth}{!}{%
\begin{tabular}{c c| c c c | c | c }
Atom & Group & {OL3} & {Shaw}  & {CHARMM36}  & {$\left<\right.$Hirshfeld$\left.\right>$ } & {$\left<\right.$xDRESP$\left.\right>$ }  \\
\hline
H  & Phosphate 
    & 0.3129 & 0.3129 & 0.34 & 0.196 & 0.457 \\
\hline
O  & Phosphate 
    & -0.621 & -0.621 & -0.68 & -0.287 & -0.528 \\
\hline
P  & Phosphate 
    & 1.1662 & 1.1662 & 1.50 & 0.372 & 0.500  \\
\hline
OP  & Phosphate 
    & -0.776 & -0.776 & -0.82 & \makecell{-0.585\\-0.554} & \makecell{-0.700\\-0.690}  \\
\hline
O5'  & Phosphate 
    & -0.4989 & -0.4989 & -0.62 & -0.192 & -0.175 \\
\hline
C5'  & Ribose 
    & 0.0558 & 0.0558 & -0.08 & 0.004 & 0.043  \\
\hline
H5'  & Ribose 
    & 0.0679 & 0.0679 & 0.09 & \makecell{0.041\\0.043} & \makecell{0.073\\0.100} \\
\hline
C4'  & Ribose 
    & 0.1065 & 0.1065 & 0.16 & 0.040 & 0.030  \\
\hline
H4'  & Ribose 
    & 0.1174 & 0.1174 & 0.09 & 0.045 & 0.123  \\
\hline
O  & Ribose 
    & -0.3548 & -0.3548 & -0.50 & -0.152 & -0.277  \\
\hline
C1'  & Ribose 
    & 0.0191 & 0.0191 & 0.16 & 0.096 & 0.084  \\
\hline
H1'  & Ribose 
    & 0.2006 & 0.2006 & 0.09 & 0.057 & 0.104 \\
\hline
C3'  & Ribose 
    & 0.2022 & 0.2022 & 0.14 & 0.032 & 0.010 \\   
\hline
H3'  & Ribose 
    & 0.0615 & 0.0615 & 0.09 & 0.041 & 0.159  \\   
\hline
C2'  & Ribose 
    & 0.0670 & 0.0670 & 0.14 & 0.036 & 0.024  \\   
\hline
H2'  & Ribose 
    & 0.0972 & 0.0972 & 0.09 & 0.049 & 0.181  \\   
\hline
O2'  & Ribose 
    & -0.6139 & -0.6139 & -0.66 & -0.250 & -0.598  \\  
\hline
HO2'  & Ribose 
    & 0.4186 & 0.4186 & 0.43 & 0.202 & 0.439  \\  
\hline
O3'  & Ribose 
    & -0.6541 & -0.6541 & -0.66 & -0.259 & -0.627  \\  
\hline
HO3'  & Ribose 
    & 0.4376 & 0.4376 & 0.43 & 0.196 & 0.451   \\  
\hline
N9  & Guanine 
    & 0.0492 & 0.0492 & -0.02 & -0.016 &  0.115  \\
\hline
C8  & Guanine 
    & 0.1374 & 0.1374 & 0.25 & 0.051 & 0.121  \\
\hline
H8  & Guanine 
    & 0.1640 & 0.1640 & 0.16 & 0.058 & 0.088  \\
\hline
N7  & Guanine 
    & -0.5709 & -0.5709 & -0.60 & -0.277 & -0.614  \\
\hline
C5  & Guanine 
    & 0.1744 & 0.1744 & 0.00 & -0.032 & 0.147  \\
\hline
C6  & Guanine 
    & 0.477 & 0.477 & 0.54 & 0.124 & 0.247 \\
\hline
O6  & Guanine 
    & -0.5597 & -0.5597 & -0.51 & -0.441 & -0.645  \\
\hline
N1  & Guanine 
    & -0.4787 & -0.5606 & -0.34 & -0.066  & -0.041  \\
\hline
H1  & Guanine 
    & 0.3424 & 0.4243 & 0.26 & 0.177 & 0.303  \\
\hline
C2  & Guanine 
    & 0.7657 & 0.7657 & 0.75 & 0.182 & 0.112  \\   
\hline
N2  & Guanine 
    & -0.9672 & -1.0158 & -0.68 & -0.144 & -0.396  \\   
\hline
H2  & Guanine 
    & 0.4364 & 0.4607 & \makecell{0.32\\0.35} & \makecell{0.174\\0.170} & \makecell{0.309\\0.317}   \\   
\hline
N3  & Guanine 
    & -0.6323 & -0.6323 & -0.74 & -0.220 & -0.448  \\   
\hline
C4  & Guanine 
    & 0.1222 & 0.1222 & 0.26 & 0.089 & 0.206  \\   
\hline
\end{tabular}
}
\caption{Point charges for Gua nucleotide from common biomolecular force fields and xDRESP fit.
Atom names refer to the scheme in Fig.~\ref{fig:SI_gua_dresp_charges}.
The values of  $\left<\right.$Hirshfeld$\left.\right>$  and $\left<\right.$xDRESP$\left.\right>$ correspond to the average during MD.}
\label{tab:SI_gua_point_charges}
\end{table}

\begin{figure}
    \centering
    \includegraphics[width=.78\linewidth]{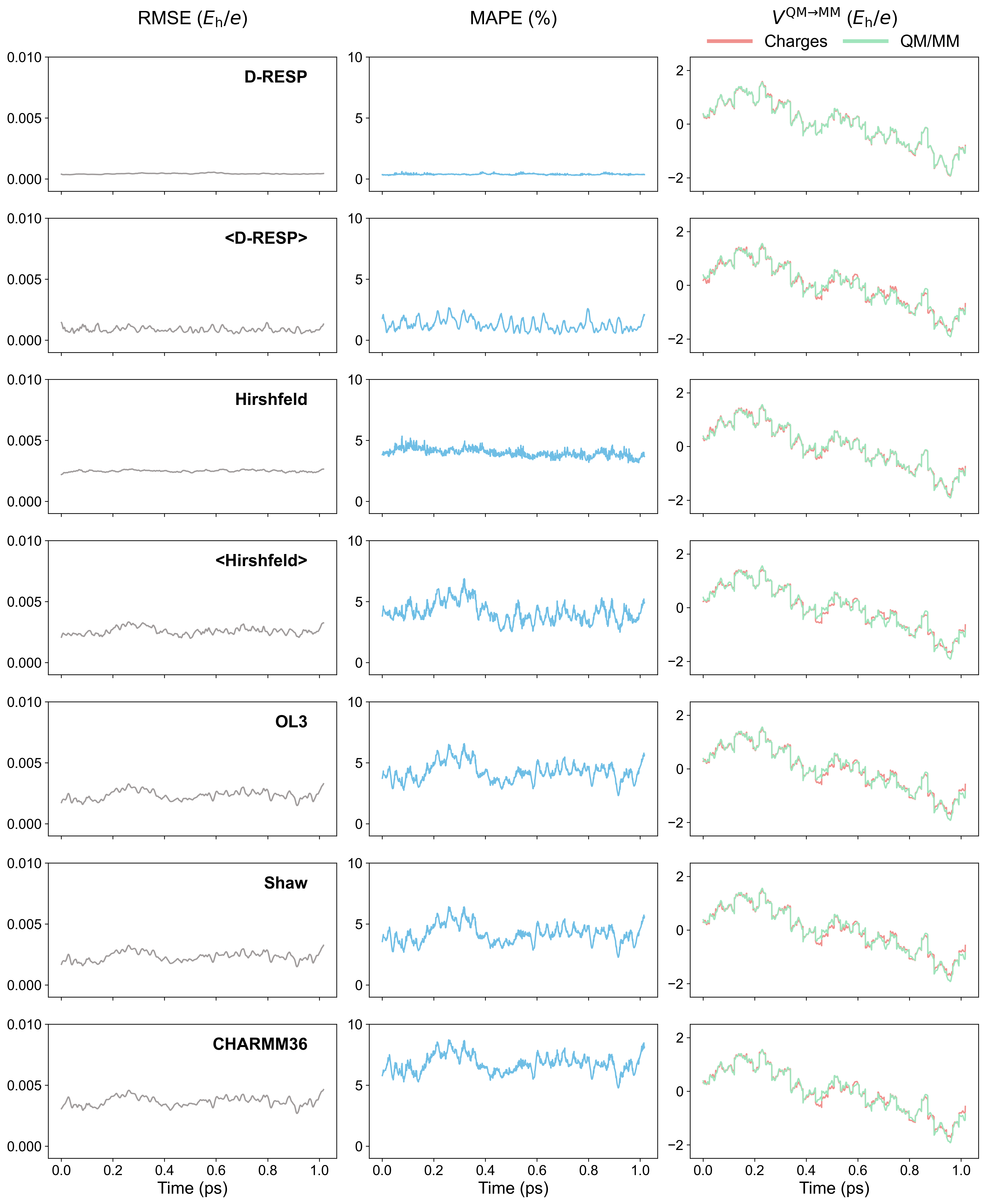}
    \caption{Comparison between xDRESP charges fitted at every MD step and different point charge models in reproducing $V^{\mathrm{QM \to MM}}$ for the Gua nucleotide system. 
    The potential is given relative to the average potential ($\Delta V^{\mathrm{QM \to MM}}(t) = V^{\mathrm{QM \to MM}}(t) - \langle V^{\mathrm{QM \to MM}}\rangle$).}
    \label{fig:SI_gua_metric_potential}
\end{figure}

\begin{figure}
    \centering
    \includegraphics[width=.78\linewidth]{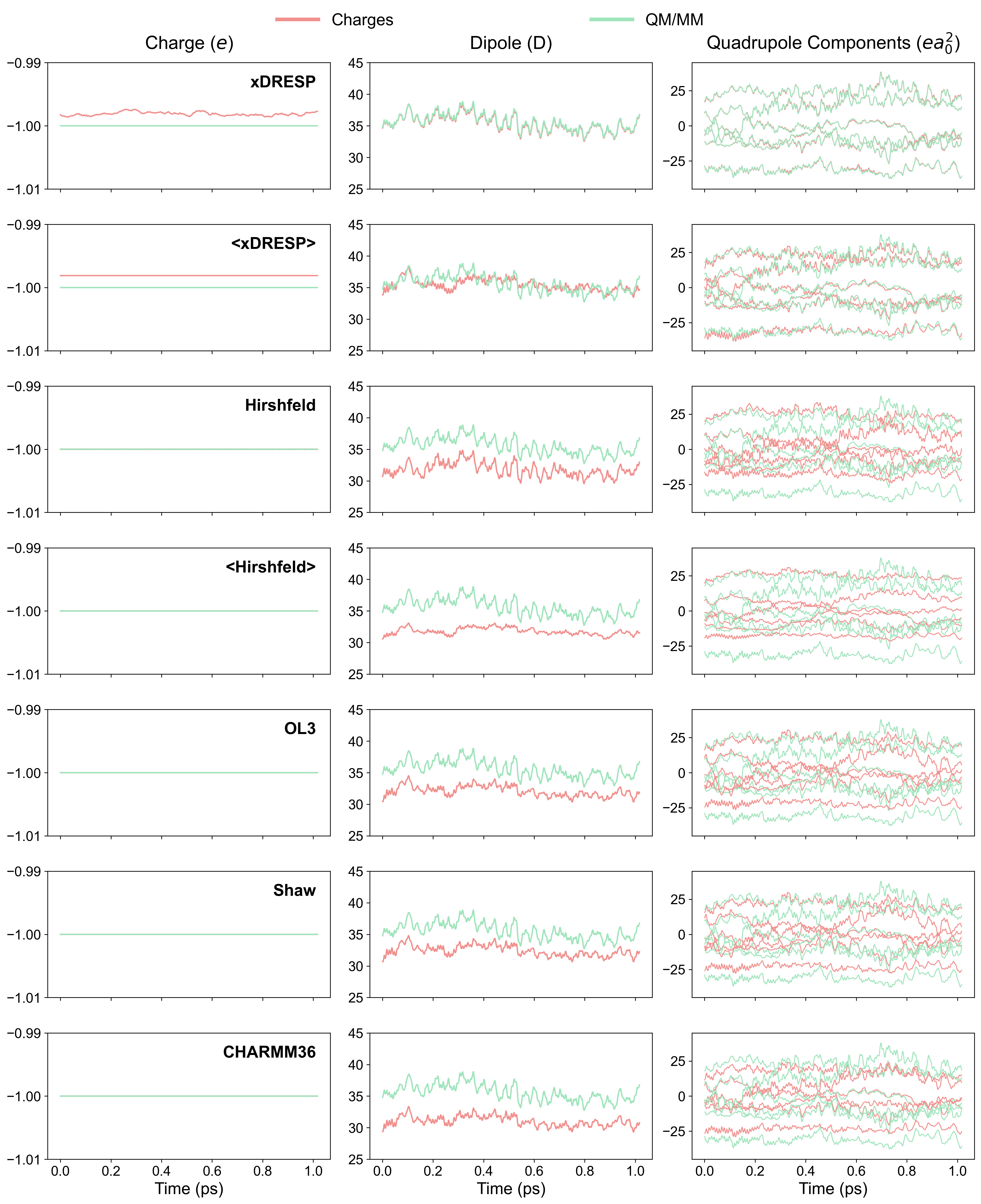}
    \caption{Comparison between xDRESP charges fitted at every MD step and different point charge models in reproducing the molecular multipoles from QM/MM for the Gua nucleotide system. 
    When not visible, the line corresponding to the point charge set (red) is overlapped by the QM/MM reference (green).}
    \label{fig:SI_gua_metric_multipoles}
\end{figure}

\newpage
\subsubsection*{CREB--APAP system}
\begin{figure}
    \centering
    \includegraphics[width=.75\linewidth]{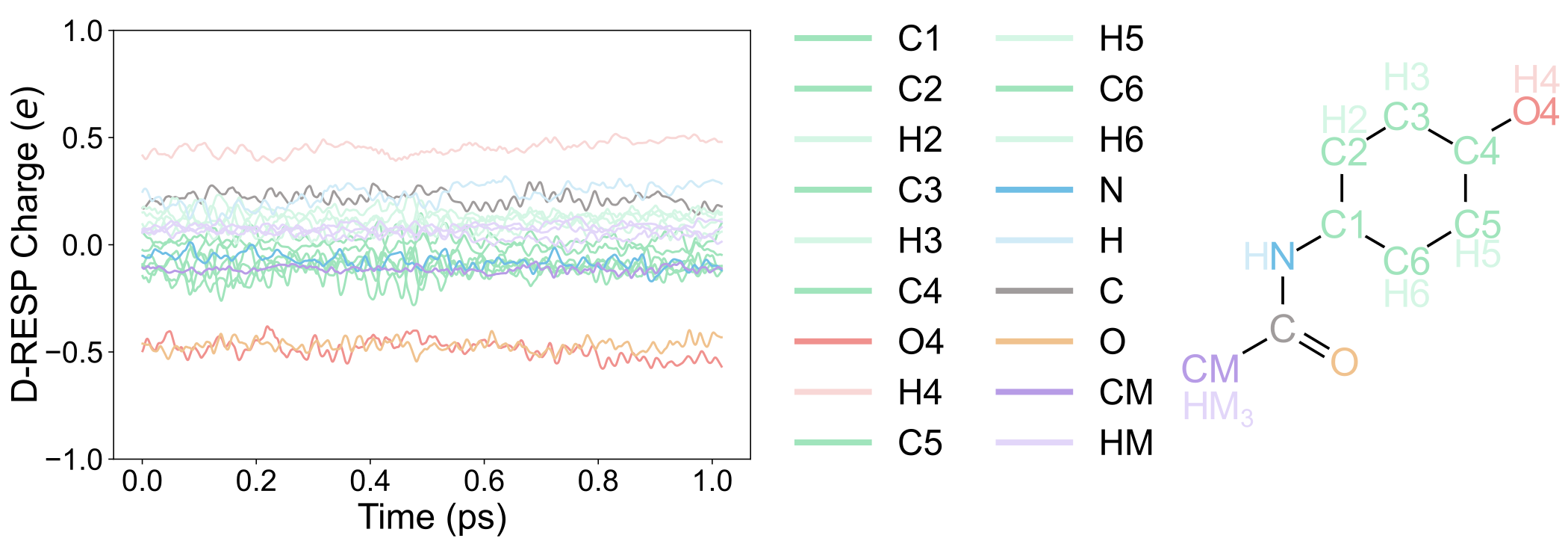}
    \caption{xDRESP charges for the APAP ligand in the CREB--APAP system dipeptide during \SI{1}{\pico\second} QM/MM MD with a restraint to the Hirshfeld charges $w_\textrm{R}$=\num{e-3}.}
    \label{fig:SI_crepapap_dresp}
\end{figure}

\begin{table}
\centering
\resizebox{\textwidth}{!}{%
\begin{tabular}{c | c c c c c c | c | c }
Atom & {RESP} & \makecell{AM1\\BCC}  & {ABCG2} & {CGenFF}  & \makecell{1.14*CM1A}  & \makecell{1.14*CM1A\\LBB}  & {$\left<\right.$Hirshfeld$\left.\right>$ } & {$\left<\right.$xDRESP$\left.\right>$ }  \\
\hline
C1 & 0.184506 & 0.037600 & -0.062000 & 0.140 & 0.2075 & 0.2075 & 0.041 & -0.008  \\
C2 & -0.166540 & -0.135500 & -0.100500 & -0.115 & -0.1283 & -0.1183 &  -0.053 & -0.105  \\
H2 & 0.169642 & 0.153000 & 0.126000 & 0.115 & 0.1596   & 0.1496 & 0.026 & 0.071  \\
C3 & -0.339418 & -0.158500 & -0.153000 & -0.115 & -0.1774 & -0.1674 & -0.065 & -0.090  \\
H3 & 0.195092 & 0.143500 & 0.121500 & 0.115 & 0.1508 & 0.1408 & 0.053 & 0.150  \\
C4 & 0.404596 & 0.096100 & 0.119100 & 0.110 & 0.1085 & 0.3285 & 0.068 & 0.022 \\
O4 & -0.533578 & -0.497100 & -0.534100 & -0.540 & -0.4974 & -0.7174 & -0.212 & -0.489  \\
H4 & 0.361069 & 0.416000 & 0.464000 & 0.430 & 0.4325 & 0.4325 & 0.215 & 0.453  \\
C5 & -0.339418 & -0.158500 & -0.153000 & -0.115 & -0.1774 & -0.1674 & -0.070 & -0.104  \\
H5 & 0.195092 & 0.143500 & 0.121500 & 0.115 & 0.1508 & 0.1408 & 0.051 & 0.114  \\
C6 & -0.166540 & -0.135500 & -0.100500 & -0.115 & -0.1283 & -0.1183 & -0.070 & -0.132  \\
H6 & 0.169642 & 0.153000 & 0.126000 & 0.115 & 0.1596 & 0.1496 & 0.055 & 0.127  \\
N & -0.538887 & -0.473100 & -0.153000 & -0.470 & -0.9606 & -0.9606 & -0.082 & -0.075  \\
H & 0.252912 & 0.312500 & 0.148000 & 0.330 & 0.4842 & 0.4842 & 0.147 & 0.232  \\
C & 0.894089 & 0.665100 & 0.606100 & 0.520 & 0.5697 & 0.5697 & 0.169 & 0.218  \\
O & -0.586809 & -0.591100 & -0.583100 & -0.520 & -0.4389 & -0.4389 & -0.336 & -0.470  \\
CM & -0.770847 & -0.179100 & -0.183100 & -0.270 & -0.2431 & -0.2431 & -0.112 & -0.115 \\
HM1 & 0.205132 & 0.068700 & 0.063033 & 0.090 & 0.1093 & 0.1093 & 0.058 & 0.074  \\
HM2 & 0.205132 & 0.068700 & 0.063033 & 0.090 & 0.1093 & 0.1093 & 0.052 & 0.046  \\
HM3 & 0.205132 & 0.068700 & 0.063033 & 0.090 & 0.1093 & 0.1093 & 0.064 & 0.081 \\
\end{tabular}
}
\caption{Point charges for the APAP ligand in the CREB-APAP system from common general force fields and xDRESP fit.
Atom names refer to the scheme in Fig.~\ref{fig:SI_crepapap_dresp}.
The values of  $\left<\right.$Hirshfeld$\left.\right>$  and $\left<\right.$xDRESP$\left.\right>$ correspond to 
the average during the MD trajectory.}
\label{tab:SI_apap_point_charges}
\end{table}

\begin{figure}
    \centering
    \includegraphics[width=.7\linewidth]{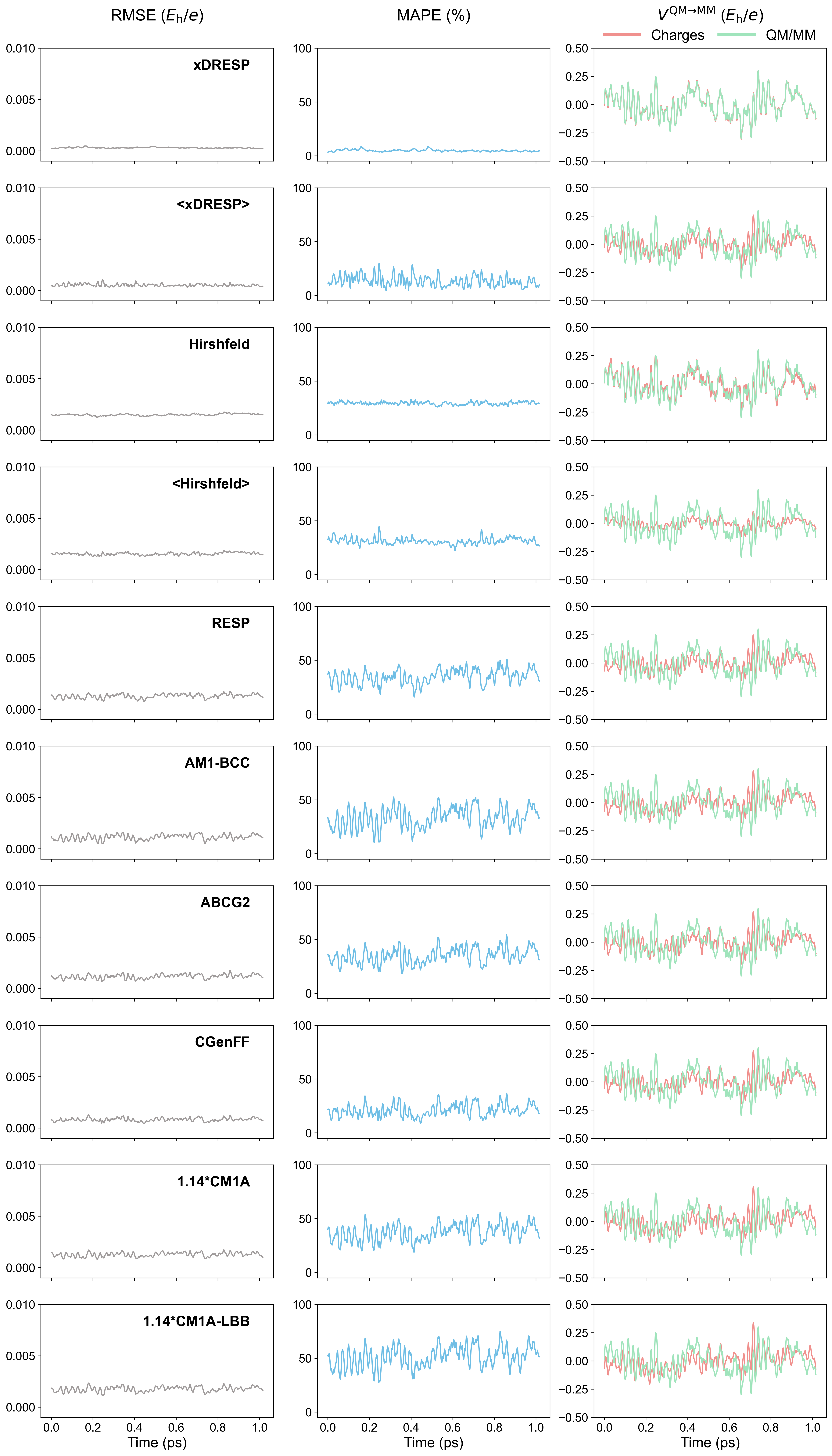}
    \caption{Comparison between xDRESP charges fitted at every MD step and different point charge models in reproducing $V^{\mathrm{QM \to MM}}$ for the APAP ligand in the CREB-APAP system. 
    The potential is given relative to the average potential ($\Delta V^{\mathrm{QM \to MM}}(t) = V^{\mathrm{QM \to MM}}(t) - \langle V^{\mathrm{QM \to MM}}\rangle$).}
    \label{fig:SI_crepapap_metric_potential}
\end{figure}

\begin{figure}
    \centering
    \includegraphics[width=.7\linewidth]{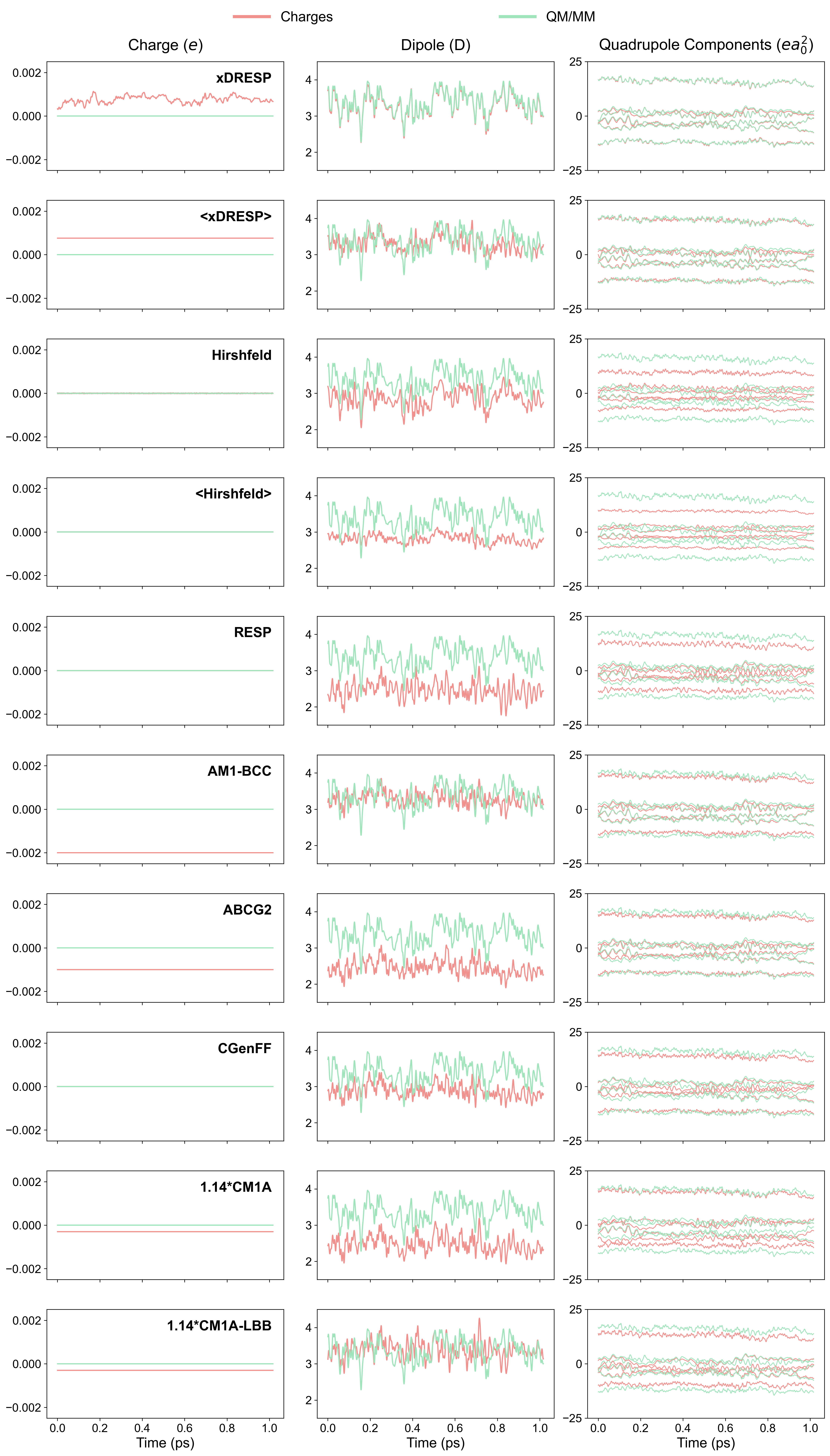}
    \caption{Comparison between xDRESP charges fitted at every MD step and different point charge models in reproducing the molecular multipoles from QM/MM for the APAP ligand in the CREB-APAP system. 
    When not visible, the line corresponding to the point charge set (red) is overlapped by the QM/MM reference (green).}
    \label{fig:SI_crepapap_metric_multipoles}
\end{figure}

\newpage
\subsection*{Effect of higher-order multipoles}
\subsubsection*{Ph--Br system}
\begin{figure}
    \centering
    \includegraphics[width=.6\linewidth]{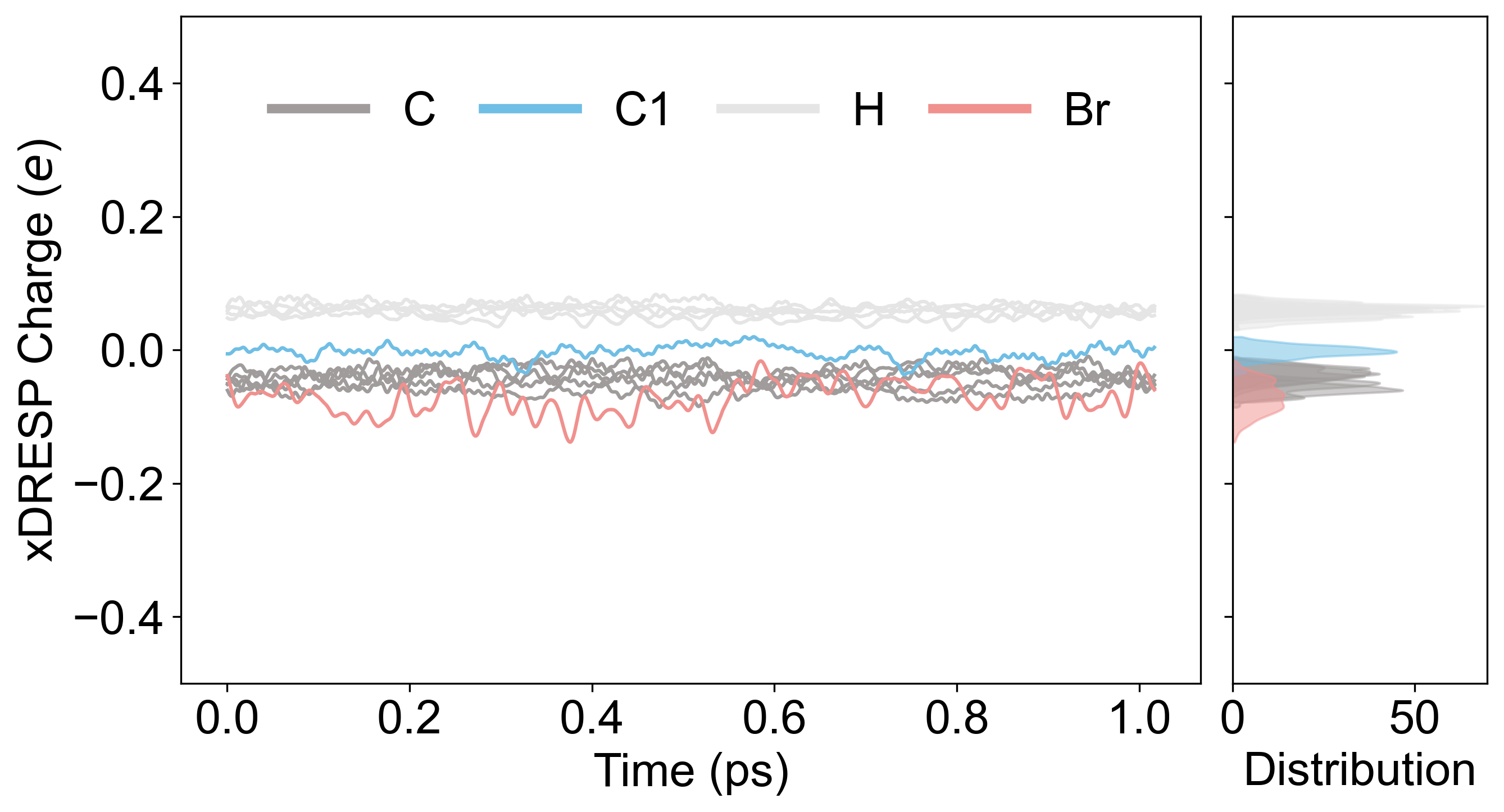}
    \caption{xDRESP charges for the Phe--Br molecule during \SI{1}{\pico\second} QM/MM MD with a restraint to the Hirshfeld charges $w_\textrm{R}$=\num{e-3}, fitting atomic charges and dipoles.}
    \label{fig:SI_dresp_Brbnz_dipole_fit}
\end{figure}

\begin{figure}
    \centering
    \includegraphics[width=.6\linewidth]{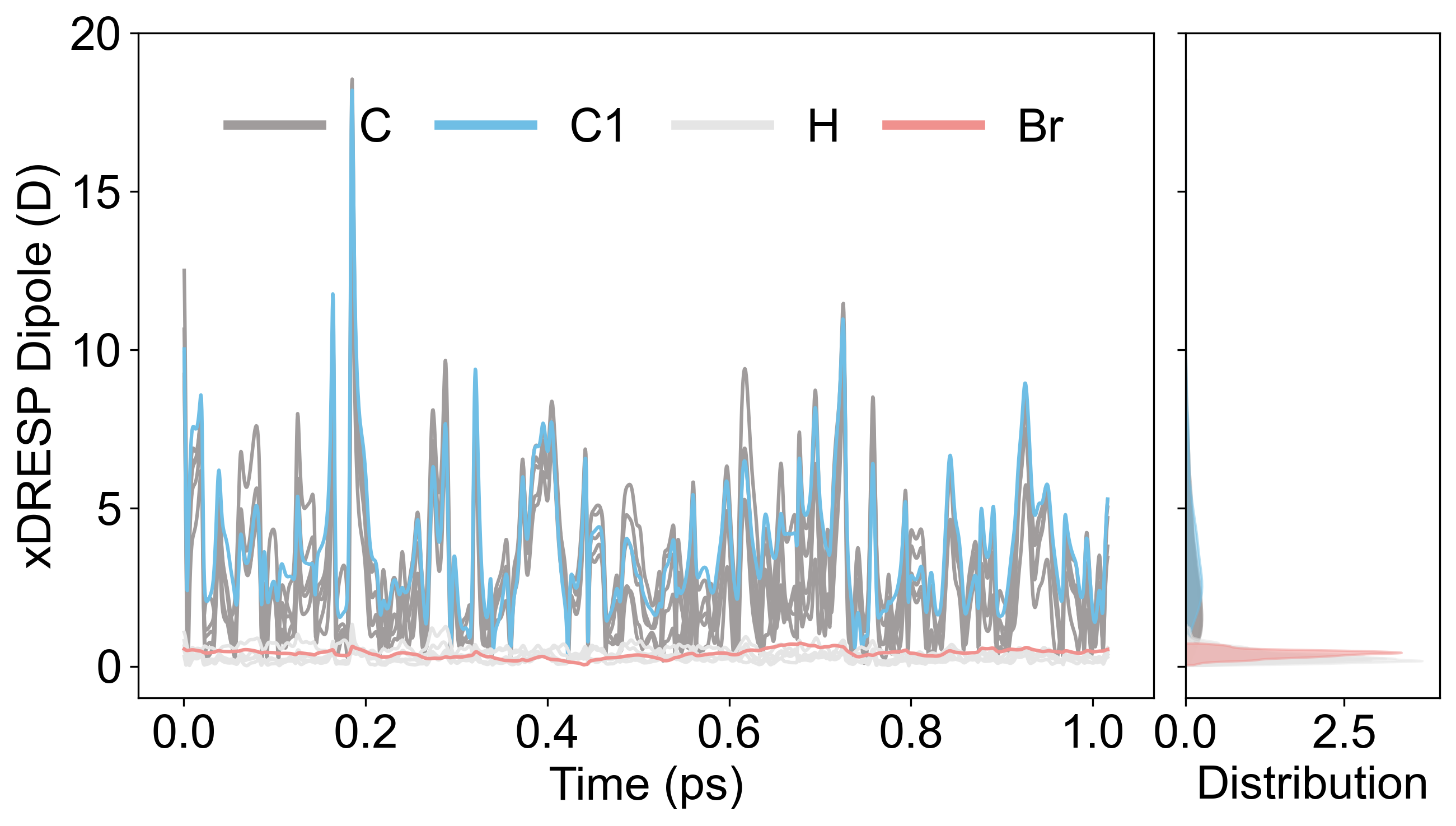}
    \caption{xDRESP atomic dipoles for the Phe--Br molecule during \SI{1}{\pico\second} QM/MM MD with a restraint to the Hirshfeld charges $w_\textrm{R}$=\num{e-3}, fitting atomic charges and dipoles.}
    \label{fig:SI_dresp_Brbnz_dipole_fit_dipole}
\end{figure}

\begin{figure}
    \centering
    \includegraphics[width=.7\linewidth]{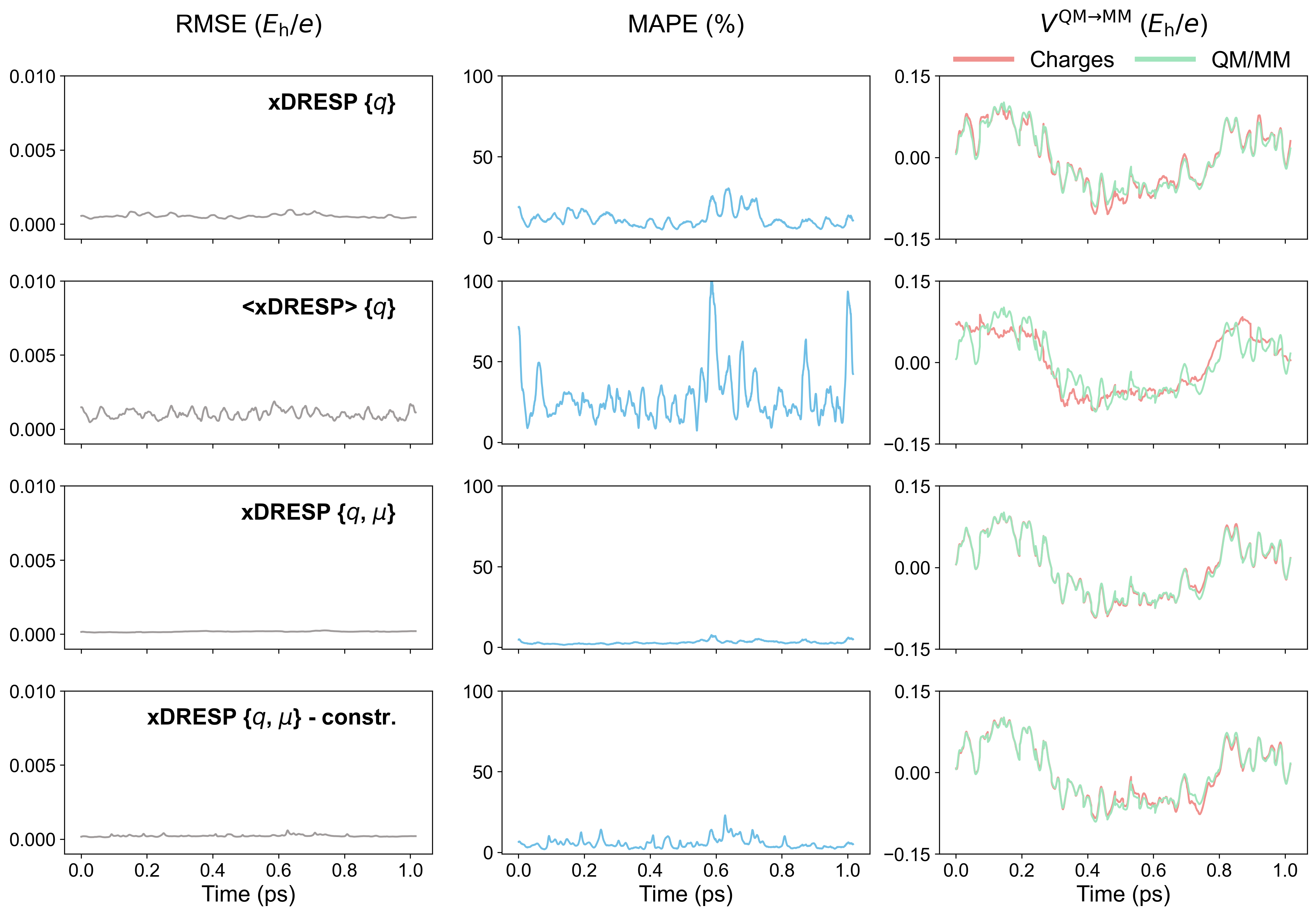}
    \caption{Comparison in the accuracy in reproducing $V^{\mathrm{QM \to MM}}$ for the Phe--Br molecule between xDRESP charges fitted at every MD step (xDRESP \{$q$\}), the fixed point charge model obtained as their average ($\left<\right.$xDRESP$\left.\right>$ \{$q$\}), and xDRESP charges and dipoles fitted at every step with no constraint (xDRESP \{$q$, $\mu$\}), and imposing constraints on the molecular charge and dipole (xDRESP \{$q$, $\mu$\} - constr.).
    The potential is given relative to the average potential ($\Delta V^{\mathrm{QM \to MM}}(t) = V^{\mathrm{QM \to MM}}(t) - \langle V^{\mathrm{QM \to MM}}\rangle$).}
    \label{fig:SI_Brbnz_metric_potential_complete}
\end{figure}

\begin{figure}
    \centering
    \includegraphics[width=.7\linewidth]{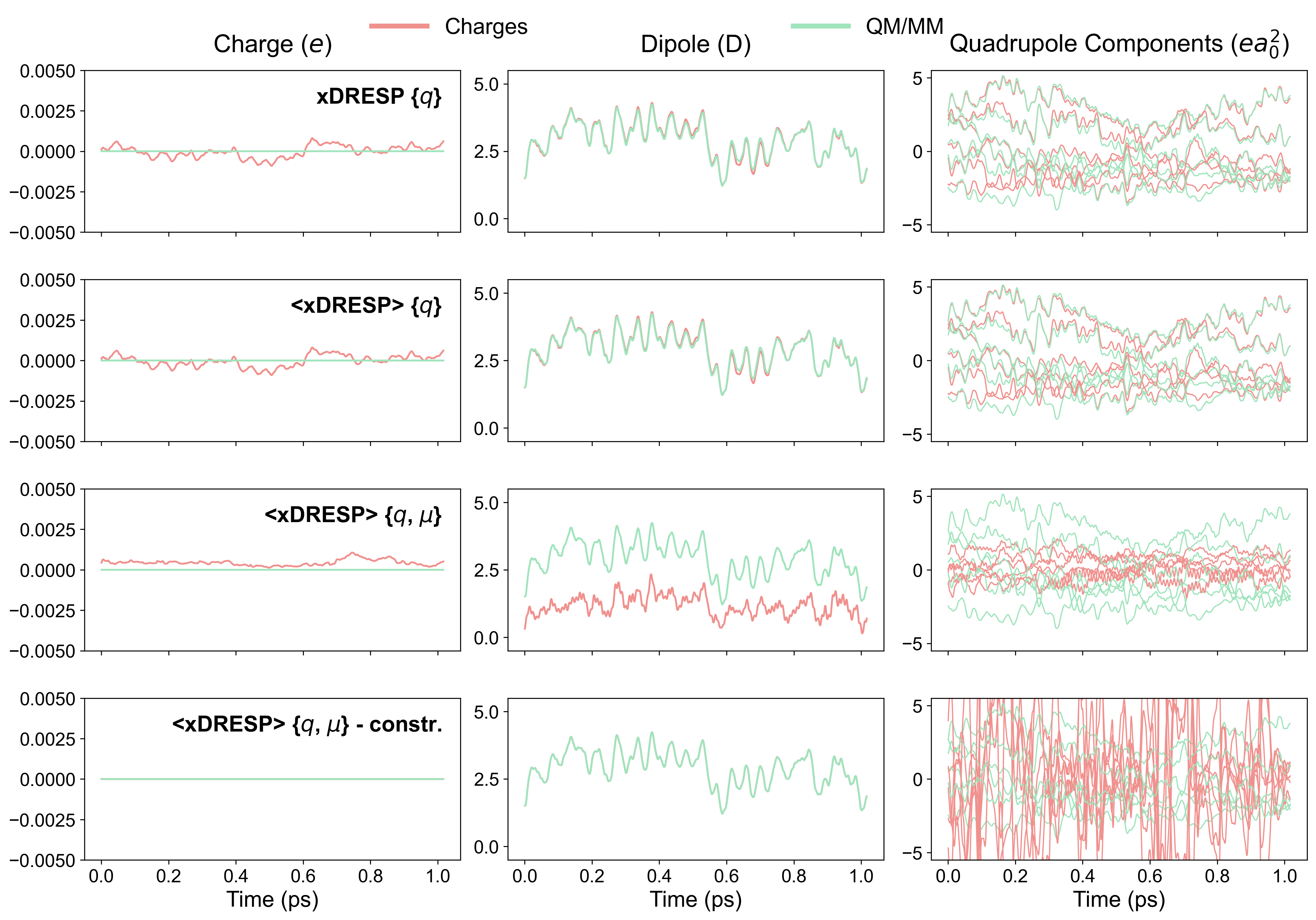}
    \caption{Comparison between xDRESP multipoles fitted at every MD step and a fixed point charge model in reproducing the molecular multipoles from QM/MM for the Phe--Br molecule. 
    When not visible, the line corresponding to the point charge set (red) is overlapped by the QM/MM reference (green).}
    \label{fig:SI_Brbnz_metric_multipoles_complete}
\end{figure}

\null
\clearpage
\subsection*{On-the-fly tracking of changes in the charge distribution}
\subsubsection*{\texorpdfstring{S$_\mathrm{N}$2}{SN2} system}

\begin{figure}
    \centering
    \includegraphics[width=.45\linewidth]{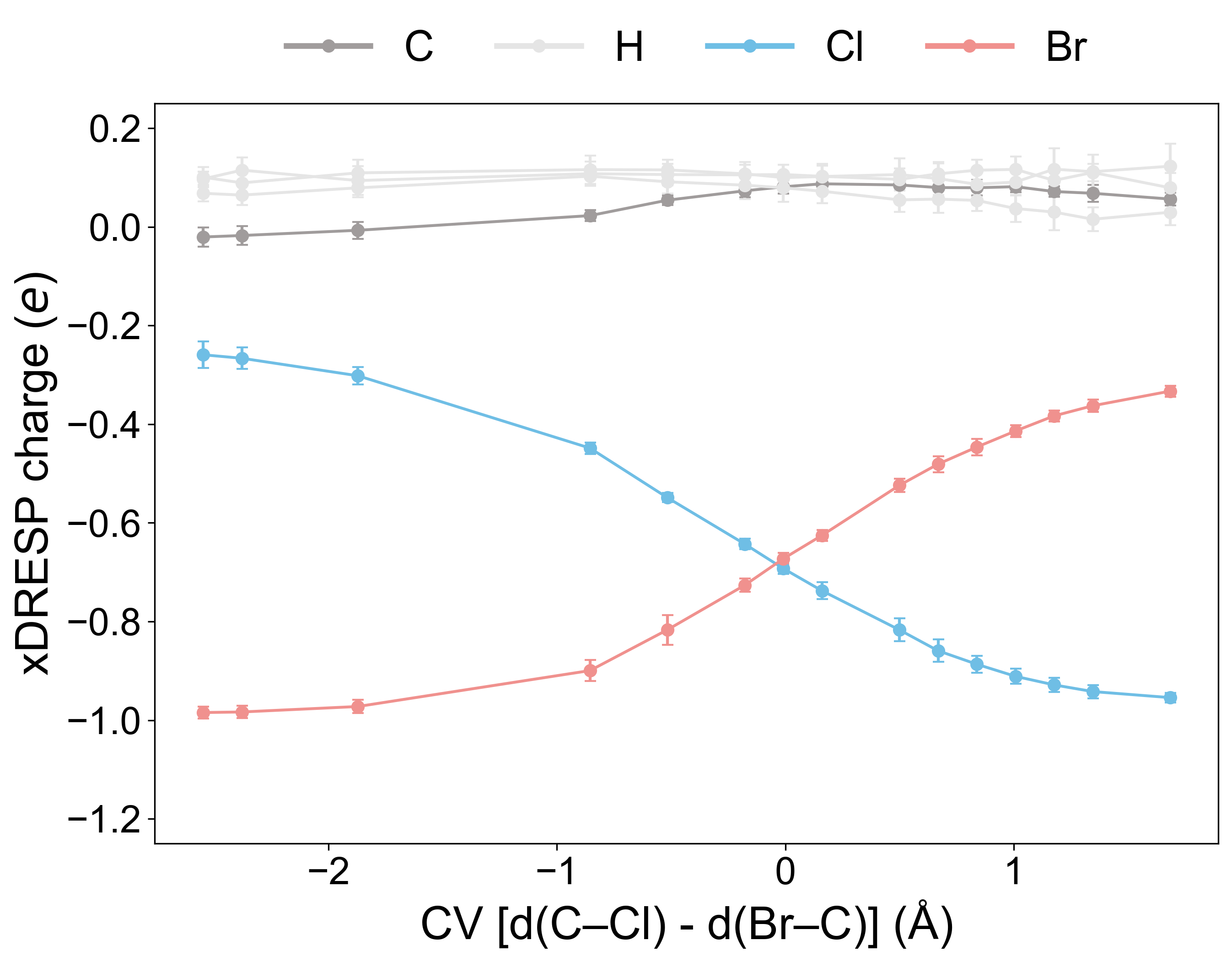}
    \caption{xDRESP charges for the S$_\mathrm{N}$2 system during thermodynamic integration, where the xDRESP fit is only performed up to zero-th order (point charges).}
    \label{fig:SI_sn2_dresp}
\end{figure}

\begin{figure}
    \centering
   \includegraphics[width=.65\linewidth]{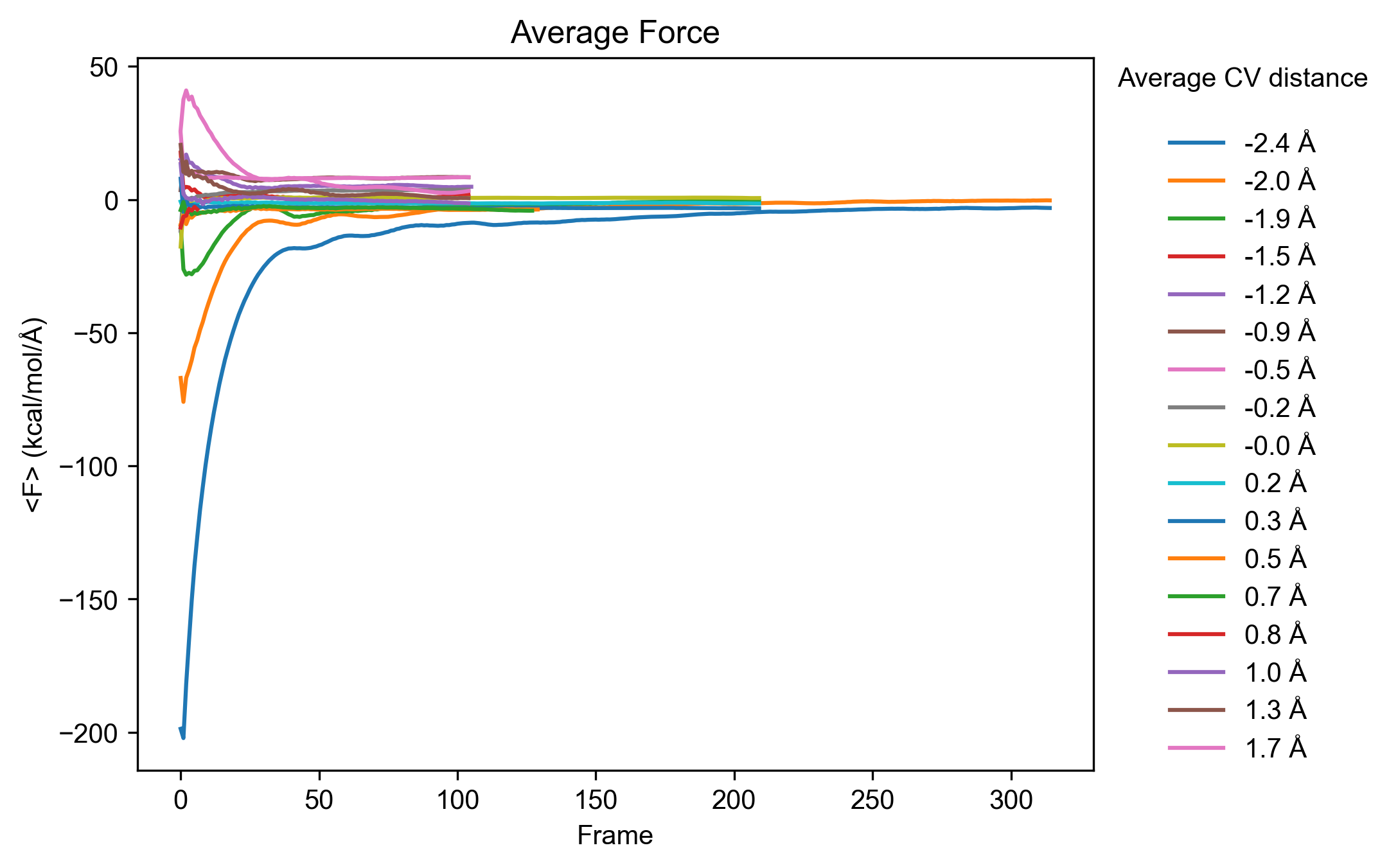}
    \caption{Convergence of the average constraint force during thermodynamic integration, where the xDRESP fit is performed up to first order (point charges and dipoles). 
The constraint force along the constrained distance along the CV (d(C--Cl) - d(Br--C) is stored every 10 MD steps, i.e., every  \num{100}\,$\hbar / E_\mathrm{h}$ ($\sim$ \qty{2.5}{\femto\second})}.
    \label{fig:SI_sn2_force_convergence_TI}
\end{figure}

\begin{figure}
    \centering
    \includegraphics[width=\linewidth]{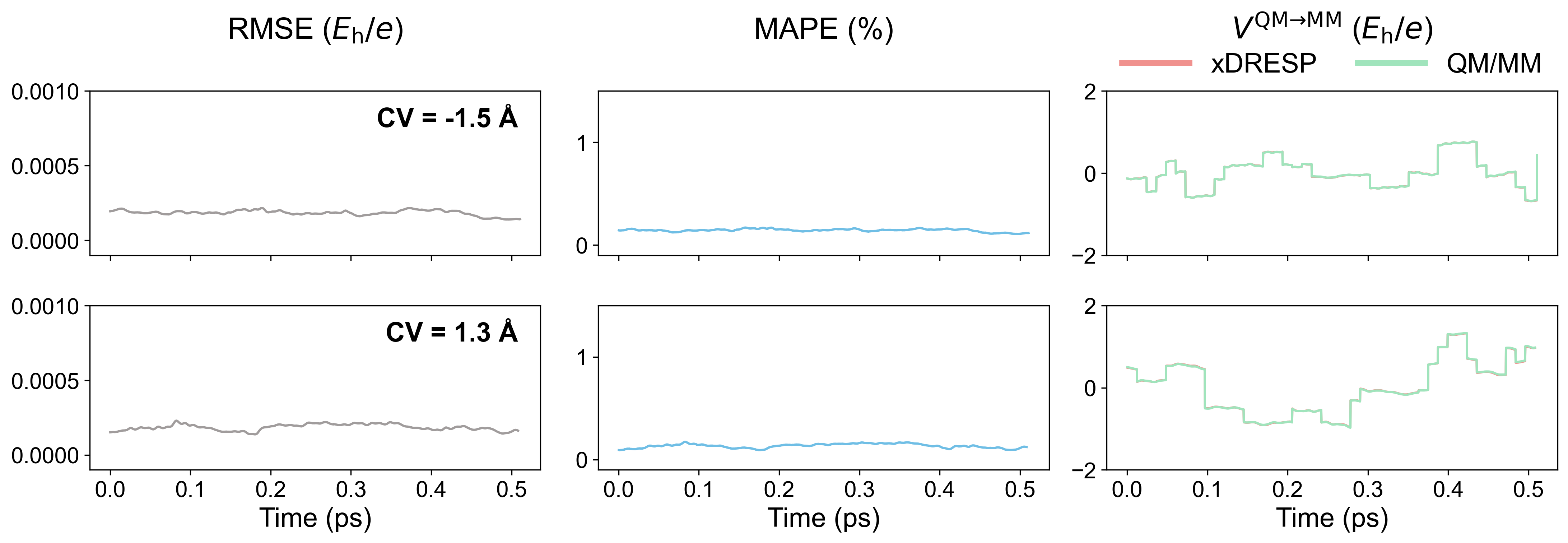}
    \caption{Accuracy of the xDRESP multipoles in reproducing $V^{\mathrm{QM \to MM}}$ during the S$_\mathrm{N}$2 reaction TI. 
    The two TI windows selected correspond to the two minima in the free energy profile.
    When not visible, the line corresponding to the xDRESP multipoles (red) is overlapped by the QM/MM reference (green).
    }
    \label{fig:SI_sn2_potential}
\end{figure}

\begin{figure}
    \centering
    \includegraphics[width=\linewidth]{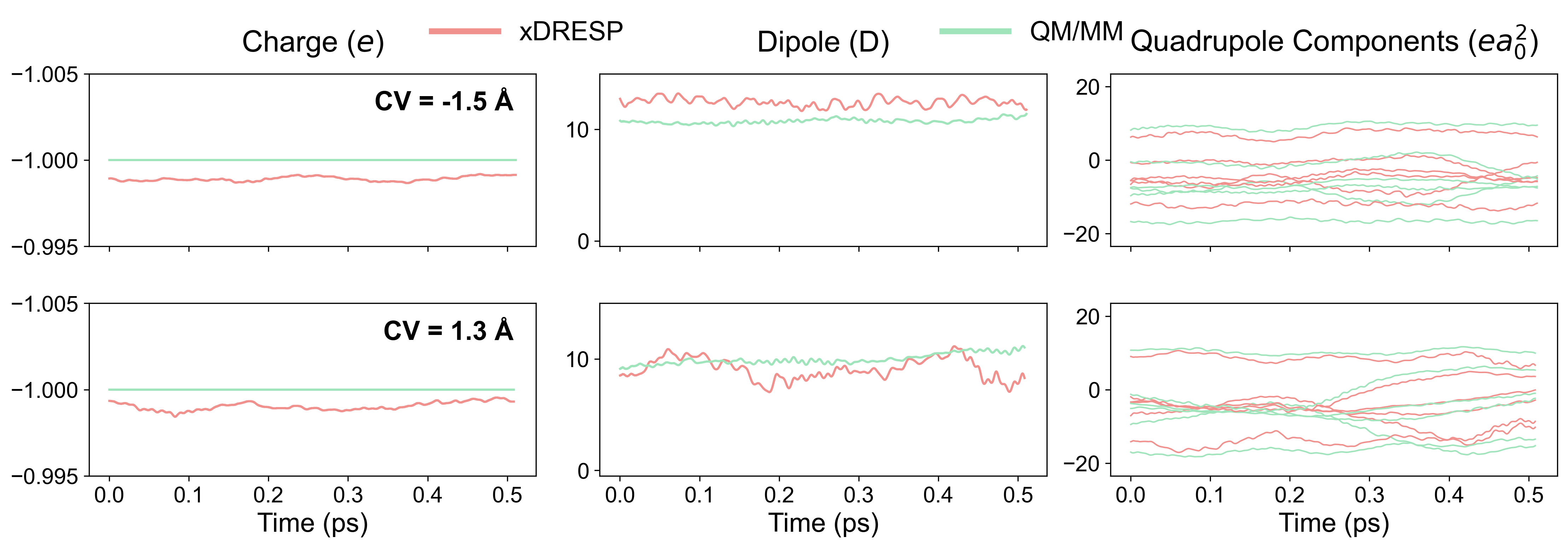}
    \caption{Accuracy of the xDRESP multipoles in reproducing the molecular multipoles for the QM subsystem during the S$_\mathrm{N}$2 reaction TI. 
    The two TI windows selected correspond to the two minima in the free energy profile.}
    \label{fig:SI_sn2_multipoles}
\end{figure}

\null
\clearpage



\end{document}